\documentclass[11pt]{article}
 \usepackage{jheppub}

\usepackage{color}
\usepackage{amsmath}
\usepackage{verbatim}
\usepackage{subfigure}
\usepackage{acronym}

\usepackage{amsfonts}
\usepackage{amssymb}
\usepackage{mathrsfs}
\usepackage{graphicx}
\usepackage{multirow}
 \usepackage{slashed}
 \usepackage{epsfig,multicol,bbm}
 \usepackage{url}

\definecolor{commam}{rgb}{0.2,0.5,1.0}
\definecolor{myred}{rgb}{0.9, 0.5, 0.1}
\definecolor{myblue}{rgb}{0, 0, 0.7}
\definecolor{mygreen}{rgb}{0.04, 0.7, 0.5}

\hypersetup{colorlinks,citecolor=myred,linkcolor=myblue,urlcolor=myblue,linktocpage=true}

 \def\be   {\begin{equation}}   \def\ee   {\end{equation}}
 \def\ba   {\begin{array}}      \def\ea   {\end{array}}
 \def\bea  {\begin{eqnarray}}   \def\eea  {\end{eqnarray}}
 \def\bean {\begin{eqnarray*}}  \def\eean {\end{eqnarray*}}
 \def\nn{\nonumber}
 \def\bry{\begin{array}}
 \def\ery{\end{array}}

\numberwithin{equation}{section}
\begin{document}

\title{Towards TeV-Scale Supersymmetric\\ Electroweak Baryogenesis}

\author[a]{Oleksii Matsedonskyi,}

\author[b,c,d]{James Unwin,}

\author[b]{and Qingyun Wang}

\affiliation[a]{DAMTP, University of Cambridge, Wilberforce Road, Cambridge, CB3 0WA, United Kingdom}

\affiliation[b]{Department of Physics, University of Illinois at Chicago, Chicago, IL 60607, USA}

\affiliation[c]{Physics Division, Lawrence Berkeley National Laboratory, Berkeley, CA 94720, USA}
\affiliation[d]{Berkeley Center for Theoretical Physics, University of California, Berkeley, CA 94720, USA}

\abstract{
Electroweak baryogenesis (EWBG) offers a compelling narrative for the generation of the baryon asymmetry, however it cannot be realised in the Standard Model, and leads  to severe experimental tensions in the Minimal Supersymmetric Standard Model (MSSM). One of the reasons for these experimental tensions is that in traditional approaches to EWBG new physics is required to enter at the electroweak phase transition, which conventionally is fixed near $\sim$100 GeV. Here we demonstrate that the addition of sub-TeV fields in supersymmetric extensions of the Standard Model permits TeV-scale strongly first-order electroweak phase transition. While earlier literature suggested no-go arguments with regards to high-temperature symmetry breaking in supersymmetric models, we show these can be evaded by employing a systematic suppression of certain thermal corrections in theories with a large number of states. The models presented push the new physics needed for EWBG to higher scales, hence presenting new parameter regions in which to realize EWBG and evade experimental tensions, however they are not expected to render EWBG completely outside of the foreseeable future experimental reach.}

\maketitle



\section{Introduction}

Symmetries and their breaking play a crucial role in modern formulation of particle physics theories. 
The symmetry structure of a theory can substantially vary during the evolution of the universe. The usual expectation is that spontaneously broken symmetries get restored at high temperatures, however the Weinberg's  work~\cite{Weinberg:1974hy} presented the prospect of high-temperature symmetry non-restoration (SNR), as well as the potential to delay symmetry restoration to higher temperatures.
SNR and related phenomena were studied in numerous subsequent papers~\cite{Mohapatra:1979qt,Fujimoto:1984hr,Salomonson:1984rh,Salomonson:1984px,Bimonte:1995xs,Dvali:1995cj,Bimonte:1995sc,Dvali:1996zr,Orloff:1996yn,Pietroni:1996zj,Gavela:1998ux,Bimonte:1999tw,Pinto:1999pg,Jansen:1998rj,Espinosa:2004pn,Aziz:2009hk,Ahriche:2010kh,Hamada:2016gux,Kilic:2015joa,Meade:2018saz,Baldes:2018nel,Glioti:2018roy,Matsedonskyi:2020mlz,Matsedonskyi:2020kuy,Chai:2020onq,Carena:2021onl,Matsedonskyi:2021hti,Bai:2021hfb,Biekotter:2021ysx,Chaudhuri:2021dsq,Biekotter:2022kgf,Agrawal:2021alq,Chang:2022psj}, with the recent applications mostly concentrating on electroweak (EW) symmetry breaking, in particular in relation to electroweak baryogenesis (EWBG)~\cite{Shaposhnikov:1987tw,Cohen:1990it} at temperatures above the EW scale. 
One of the main reasons for such an interest is that many (although, not all) EWBG models currently face an increasing pressure from the null results in searches for the new physics which is required for successful EWBG. This new physics includes new sources of CP-violation, and modifications of the Standard Model (SM) needed to make the electroweak phase transition (EWPT) strongly first-order, such that the Higgs VEV crosses the value $h\simeq T$ during the transition. 
These scenarios tend to be highly testable since the new physics introduced to permit successful EWBG  has to operate at the EWPT, which is conventionally expected to happen at temperatures around $100$~GeV, and this restricts the mass scale of new physics to be similarly light if it is to achieve its purpose.

\begin{figure}[t]
\center
\includegraphics[width=12.cm]{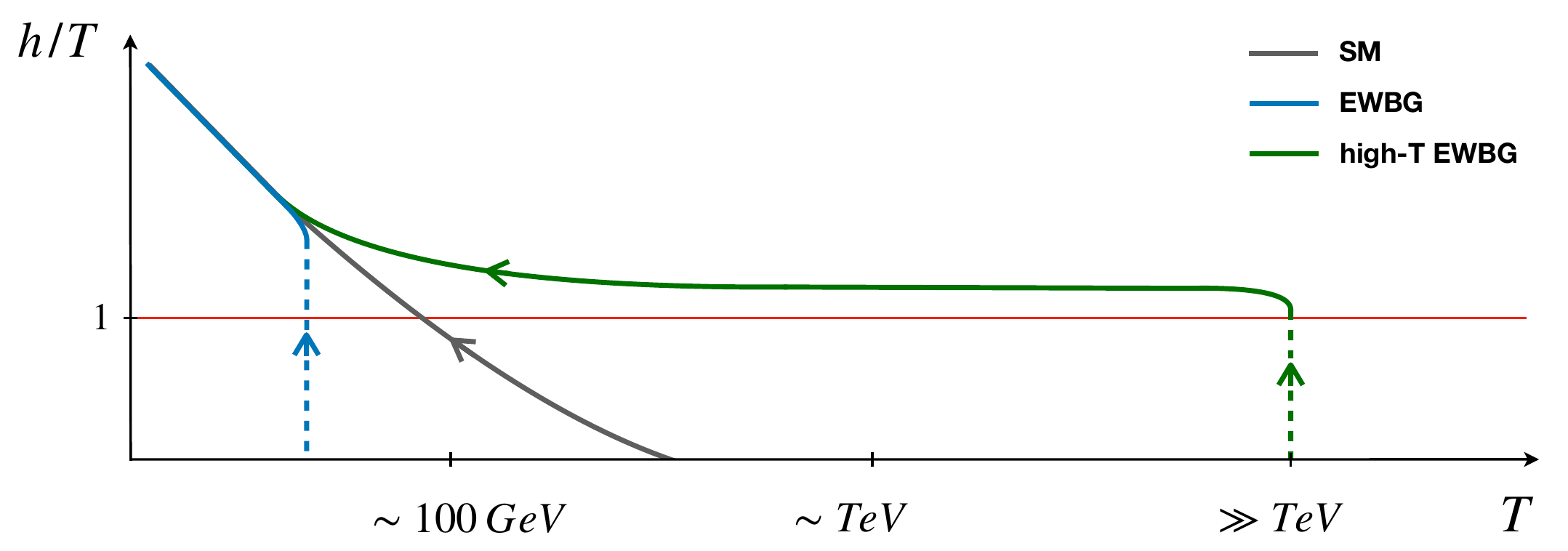}
\caption{Higgs VEV thermal histories in SM (grey), traditional models of electroweak baryogenesis (blue) and models of high-temperature EWBG (green). Dashed vertical lines correspond to first-order phase transitions. Red line shows the border between $h/T \lessgtr 1$ phases, which in EWBG models have to be separated by the first-order phase transition.}
\label{fig:general}
\end{figure}

Although traditional models of EWBG necessarily change the properties of the EW phase transition with respect to the SM predictions, the overall transition temperature stays around 100~GeV due to the thermal effects induced by the SM particles. The SM fields produce a positive thermal correction to the Higgs mass which takes over the negative zero-temperature mass at $T\sim 100$~GeV leading to symmetry restoration, see Figure~\ref{fig:general}.
The most sizeable such a correction is that of the top quark:
\be\label{eq:introtop}
{\cal L}_{\text{SM}} \supset - \lambda_{t} \bar t t H \;\; \Rightarrow \;\;  \delta V_T \sim \lambda_{t}^2 T^2 |H|^2.
\ee

The idea of high-temperature EWBG~\cite{Baldes:2018nel,Glioti:2018roy} is to raise the temperature of EW symmetry restoration, potentially to the multi-TeV range, where the phase transition would occur.  The new physics required for EWBG can then be correspondingly heavier and evade the current or near future experimental bounds. This is achieved by adding a new set of fields which induce a negative thermal Higgs mass, thus counteracting the corrections induced by the SM fields and moving the EW symmetry  restoration to higher temperatures. As an example we will consider new scalars $\chi$ with the interaction
\be
{\cal L}_{\text{\rm SNR}} \supset \lambda_{\chi h} \chi^2 |H|^2 \;\; \Rightarrow \;\;  \delta V_T \sim  - \lambda_{\chi h} T^2 |H|^2\,,
\ee
where the negative thermal Higgs mass in $\delta V_T$ has to overcompensate the positive SM contributions such as that of Eq.~(\ref{fig:general}).
Other than that, in the scenarios considered so far  the SM has not been modified while being extrapolated to the multi-TeV energies. 
We note that this approach is generically incompatible with the models of EWBG which are motivated by the gauge hierarchy problem, 
which typically predict qualitative changes to the theory, and typically new states, at energies above $\sim 1$~TeV.  Thus in this work we seek to work towards realizing the prospect of high-temperature EWBG within a supersymmetric extension of the Standard Model introduced to resolve the hierarchy problem.\footnote{An alternative solution to the hierarchy problem can be found in, for instance, Higgs compositeness at the TeV scale, with the EWBG implementation studied for example in~\cite{Bruggisser:2018mus,Bruggisser:2018mrt}.}

EWBG realizations within the minimal supersymmetric extensions of the Standard Model (such as the MSSM and NMSSM) experience severe pressure from the non-observation of new physics at collider experiments, see e.g.~\cite{ATLAS:2020syg}, and electric dipole moment (EDM) measurements~\cite{Katz:2015uja,Li:2008ez}. 
This suggests incorporating SNR in these scenarios. 
In these supersymmetric extensions the Higgs mass naturalness problem is addressed by assuming the presence of relatively light superpartners of the states which couple sizeably to the Higgs field, such as the top quark, thus cancelling their loop contributions to the Higgs mass. We highlight that models of SNR typically introduce new states $\chi$ that must couple to the Higgs with a strength  $\lambda_{\chi h}$ comparable to the top Yukawa, in order to compensate for the positive top-quark-induced thermal Higgs mass~(\ref{eq:introtop}). It follows that for SNR scenarios to comply with the EW scale naturalness considerations one must necessarily supersymmetrize the SNR sector as well.

High-temperature symmetry breaking in supersymmetric theories has been considered in previous work~\cite{Haber:1982nb,Mangano:1984dq,Bajc:1996kj,Bajc:1996id,Dvali:1996np,Dvali:1998ct,Bajc:1998jr,Bajc:1999he,Riotto:1997tf,Bajc:1998rd} in application to various problems of particle physics, finding a set of arguments preventing SNR~\cite{Haber:1982nb,Mangano:1984dq,Bajc:1996kj,Bajc:1996id}, and also proposing some ways to overcome these~\cite{Dvali:1998ct,Bajc:1998jr,Bajc:1999he,Riotto:1997tf,Bajc:1998rd}. In this paper we will show that these no-go arguments do not apply in cases with a large number of SNR states, and present a framework for EWBG utilizing this feature. 
This model, although allowing for SNR, requires a quadratically increasing number of new states for achieving higher SNR temperatures. As a result, while shifting the EWPT to the TeV scale requires a moderate ${\cal O}(10)$ number of SNR fields, achieving a phase transition at multi-TeV temperatures is much more expensive in terms of the number of new degrees of freedom. 
This behaviour represents a significant deviation from what is expected in non-supersymmetric SNR scenarios with scalar SNR fields~\cite{Meade:2018saz,Baldes:2018nel,Glioti:2018roy}, and suggests that, although the EWSB signatures at current or near future experiments can be suppressed, their complete removal is not possible unless an extremely large number of new states is postulated.  Moreover, extremely large numbers of states are not very compelling since this will generically suppress the final baryon asymmetry $\eta$ by the simple scaling argument $\eta\propto 1/g_*(T)$, where $g_*$ is the effective number of degrees of freedom \cite{Glioti:2018roy}.

The aim of this paper is to present a proof-of-principle that a supersymmetric extension of the SM can permit a strongly first-order phase transition at temperatures well above the EW scale. The model that we arrive at has a spectrum with the SM superpartners at the TeV scale or above, while the SM is supplemented with sets of scalar-fermion superpartner pairs below the TeV scale for the purpose of SNR, and heavier superpartners enhancing the phase transition to be strongly first order. This model may not be the most economic realization, but it provides the desired proof-of-principle. Furthermore, this work focuses primarily on obtaining a strongly first-order phase transition near the TeV scale and we leave to a future publication a more complete calculation of the baryon asymmetry $\eta$ in this setting. While we include some discussion of  potential sources of CP violation and the preferred particle spectra which avoid suppressions to $\eta$ in Section \ref{sec:cp}, a full calculation of $\eta$ must include a computation of the relevant CP sources, baryon number transport, and the thickness and velocity of the bubble wall associated to the phase transition, accordingly this merits a dedicated paper.

\newpage

The paper is organized as follows. In Section~\ref{sec:snrgen} we revisit the no-go arguments for SNR with SUSY and show how the large-$n_\chi$ limit helps to avoid them. In Section~\ref{sec:ewsnr} we consider in detail the application of this idea to the EW symmetry breaking. In Section~\ref{sec:ewpt} we present a simplistic model which, besides SNR, also features a high-temperature  first-order EW phase transition, as necessary for EWBG. Subsequently, in Section~\ref{sec:cp} we outline the requirements for successful EWBG, in particular we discuss potential sources of CP violation which arise within supersymmetric models. Finally, we conclude in Section~\ref{sec:disc} and the appendices contain a discussion of higher-order thermal corrections in our scenario.

\section{Symmetry Non-Restoration in Supersymmetric Theories}\label{sec:snrgen}

The possibility to have broken symmetries at high temperature in supersymmetric theories was considered in the past, in attempts to avoid the formation of monopoles or domain walls. The conclusion however was that in the simplest cases there is a series of obstacles~\cite{Haber:1982nb,Mangano:1984dq,Bajc:1996kj,Bajc:1996id} which prohibit symmetry breaking at very high $T$. We will now review these arguments and highlight a new way to overcome them.

\subsection{A `No-Go' Theorem for High Temperature Symmetry Breaking}

Consider a scalar field $\phi$ transforming non-trivially under a symmetry $G$, such that a non-zero $\phi$ vacuum expectation value (VEV) would break this symmetry spontaneously. We will analyse whether the thermal effects can drive the $\phi$ VEV to large values. 

To determine the effect of the high-temperature plasma on  $\phi$ we first write down the thermal potential in high-$T$ expansion, i.e. assuming all particles in plasma having mass $m \lesssim T$:
\be\label{eq:deltaVT1}
\delta V_T = \frac {T^2}{24} \text{Tr}\left[{\cal M}_0^2 + {\cal M}_{1/2}{\cal M}_{1/2}^{\dagger} + 3 {\cal M}_{1}^2\right],
\ee
where ${\cal M}_{i}$ are the mass matrices of particles with spin $i$, which are functions of the $\phi$ field. Assuming that supersymmetry is at most softly broken, the supertrace of mass matrices has to be independent of $\phi$
\be\label{eq:supertr}
\text{Tr}\left[{\cal M}_0^2 -2 {\cal M}_{1/2}{\cal M}_{1/2}^{\dagger} + 3 {\cal M}_{1}^2 \right] \ne f(\phi).
\ee
Using this relation we can now express the spin-0 and spin-1 mass matrices as a function of ${\cal M}_{1/2}$, to obtain
\be\label{eq:deltaVT2}
\delta V_T = \frac {T^2}{8} \text{Tr}\left[{\cal M}_{1/2}{\cal M}_{1/2}^{\dagger}\right] = \frac {T^2}{8} \sum_{i j} |{\cal M}_{1/2\,ij}|^2.
\ee
It follows that in $G$-symmetric renormalizable theories, the most general form of the sum is
\be\label{eq:summ12}
\sum_{i j} |{\cal M}_{1/2\,ij}|^2 = c_0 + (c_1 \phi + {\rm h.c.}) + c_2\, \phi^\dagger \phi,
\ee
where we omitted the $G$-indices for simplicity. The quantity on the LHS of Eq.~(\ref{eq:summ12}) is, trivially, non-negative. To ensure this at large $\phi$ field values, $c_2\, \phi^\dagger \phi$ has to be a positive semidefinite quadratic form.   
We then find that the thermal potential (\ref{eq:deltaVT2}) produces non-negative thermal masses $\propto T^2 c_2\, \phi^\dagger \phi$ and hence, if non-vanishing, drives the $\phi$ field towards the symmetry-restoring minimum.

Let us discuss the effect of a non-vanishing $c_1$ coefficient. Since all the terms on the RHS of Eq.~(\ref{eq:summ12}) have to be $G$-invariant, the coefficient $c_1$ can only be non-zero if it is proportional to some other scalar fields, which we collectively denote $\phi'$, transforming non-trivially under $G$. By the same reasoning as before we conclude that the $\phi,\phi'$-dependent part of the sum (\ref{eq:summ12}) has to be positive-semidefinite, leading to non-negative thermal masses for both fields. This leaves open a possibility to have a flat direction in $\phi-\phi'$ plane. One can generically expect that in such a situation the position of the minimum of the sum of the zero-temperature potential and its finite-$T$ correction will not be able to experience a significant growth with temperature, and hence such a scenario is also not relevant for our purposes.

The above conclusions are based on the following assumptions: 
\begin{itemize}
\item[(I)] All of the fields are light: $m_i \lesssim T$; 
\item[(II)] All of the fields are in thermal equilibrium; 
\item[(III)] None of the fields carry a net charge (together with (I) and (II) implying Eq.~(\ref{eq:deltaVT1}));  
\item[(IV)] All of the interactions are renormalizable  (implying Eq.~(\ref{eq:summ12})).  
\end{itemize}
We will now discuss the consequences of breaking each of these assumptions.
First of all, the assumption (I) is violated if some of the fields are much heavier than $T$ and hence their masses should be dropped from Eq.~(\ref{eq:deltaVT1}). On the other hand, these heavy degrees of freedom can be integrated out of the theory, giving rise to a set of non-renormalizable interactions. If the heaviness of integrated out heavy fields is achieved without introducing a naturalness problem for the $\phi$ field, which is the case we are interested in, then the property of the supertrace~(\ref{eq:supertr}) would hold for the remaining light degrees of freedom. Thus this case is  equivalent to breaking the condition (IV) which we will discuss in its turn.

Another proposed way to get SNR is to violate the assumption (II). The authors of Ref.~\cite{Bajc:1998jr} proposed to suppress the interactions of the relevant scalar field assuming the presence of flat direction in the potential. As a result the field doesn't thermalize, and hence part of the thermal effects should also be dropped from Eq.~(\ref{eq:deltaVT1}).  Furthermore, in Ref.~\cite{Riotto:1997tf} the authors showed how a non-zero net lepton number, violating the assumption (III), can lead to high-temperature symmetry breaking\footnote{An alternative approach using large net charges was recently proposed in \cite{Chang:2022psj}, which might also be applicable in the supersymmetric case.}.

\subsection{Non-renormalizable Operators and Symmetry Non-Restoration}

\begin{figure}[t]
\center
\includegraphics[width=12.cm]{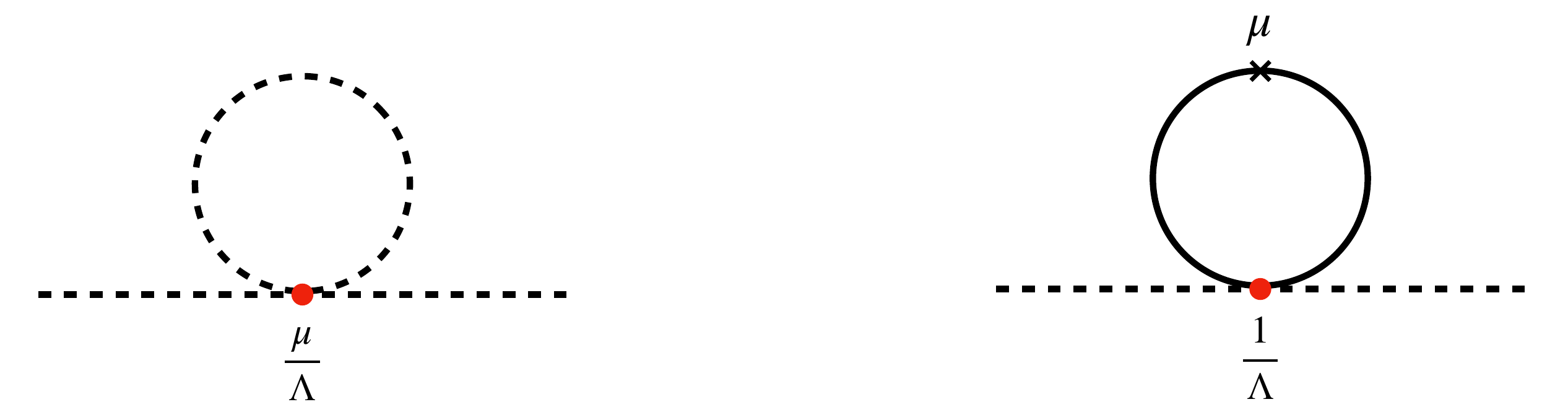}
\caption{One-loop thermal corrections to the scalar mass from the scalar (left diagram) and fermionic (right diagram) degrees of freedom in models with non-renormalizable interactions (\ref{eq:Wnonren1}), (\ref{eq:ngt1}).}
\label{fig:1_loop}
\end{figure}

Finally, the authors of Ref.~\cite{Dvali:1996np} proposed to use non-renormalizable interactions to get SNR by violating the condition (IV). In the presence of dimension-five interactions, the fermionic mass takes the form (we assume no extra scalars this time)
\be\label{eq:summ12_2}
\sum_{i j} |{\cal M}_{1/2\,ij}|^2 = c_0 + c_2\, \phi^\dagger \phi + c_4\, (\phi^\dagger \phi)^2,
\ee
hence the behaviour of the whole expression at large $\phi$ values is controlled by $c_4$, while $c_2$ is allowed to be negative, generating a negative thermal correction to the Higgs mass. 
As an example, Ref.~\cite{Dvali:1996np} considered a model with a superpotential 
\be\label{eq:Wnonren1}
W= - \frac 1 2 \mu \phi^2 + \frac 1 4 \phi^4/\Lambda,
\ee
featuring a $\phi \to - \phi$ symmetry, whose breaking at high $T$ was analysed. The corresponding scalar potential reads (we use the same notation for the superfields and their scalar components)
\bea\label{eq:nonrenormV}
V &=& |\phi|^2 |\mu - \phi^2/\Lambda|^2 \\
&=& \frac 1 2 \mu^2 (\phi_1^2+\phi_2^2) - \frac {\mu}{2\Lambda} (\phi_1^4-\phi_2^4) + \frac 1 {8\Lambda^2} (\phi_1^2+\phi_2^2)^3. \label{eq:vphiR} 
\eea
To compute the leading order thermal correction we use Eq.~(\ref{eq:deltaVT1}), and remove the fermionic masses using Eq.~(\ref{eq:supertr}), thus making $\delta V_T$ a function of scalar masses, which can be read from Eq.~(\ref{eq:nonrenormV}). We thus obtain
\be\label{eq:deltavphiR1}
\delta V_T = \frac {T^2}{16} \text{Tr}\left[{\cal M}_0^2\right] \supset \frac {3}{8}  \frac{\mu}{\Lambda} T^2 (\phi_2^2-\phi_1^2) + \frac {9 T^2}{32\Lambda^2} (\phi_1^2+\phi_2^2)^2.
\ee
Corresponding one-loop diagrams are shown in Figure~\ref{fig:1_loop}.

\begin{figure}[t]
\center
\includegraphics[width=12.cm]{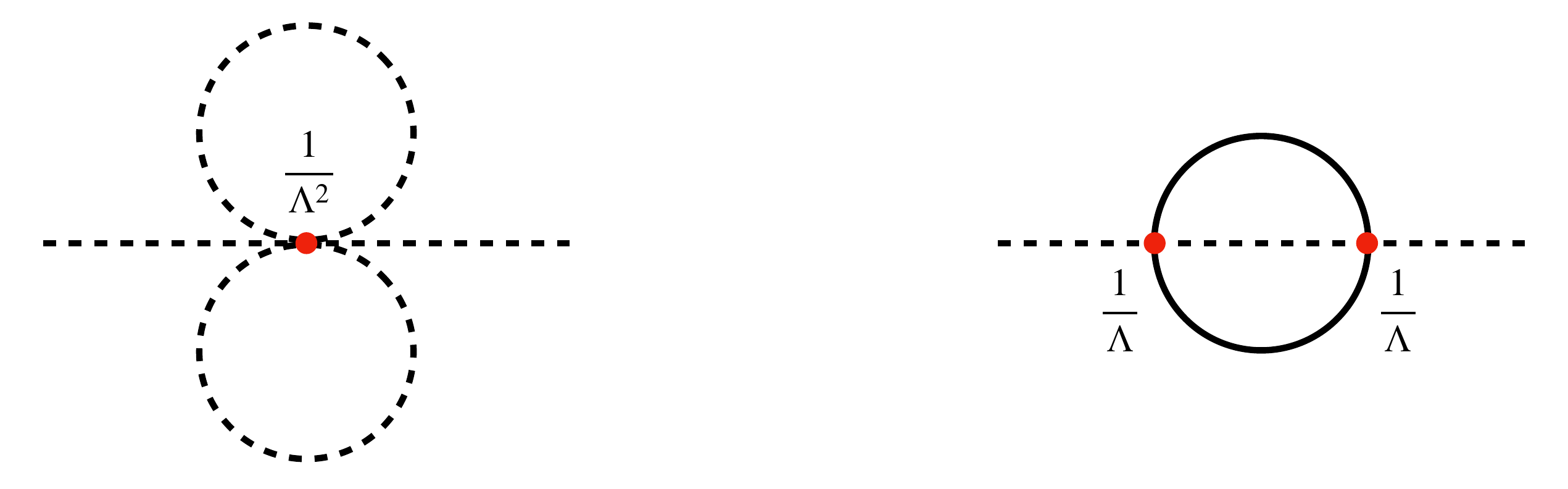}
\caption{Two-loop thermal corrections to the scalar mass from the scalar (left diagram) and fermionic (right diagram) degrees of freedom in models with nonrenormalizable interactions (\ref{eq:Wnonren1}), (\ref{eq:ngt1}).}
\label{fig:2_loop}
\end{figure}

We see from Eq.~(\ref{eq:deltavphiR1}) that the $\phi_1$ field receives a negative mass correction (we assume $\mu,\Lambda>0$), which can overcome the positive zero-temperature mass in Eq.~(\ref{eq:vphiR}) if
\be\label{eq:phiRsnr1}
T^2 > \frac 4 3 \mu \Lambda,
\ee
thus destabilizing the $\phi_1$ potential around the origin and allowing the field to get a symmetry-breaking VEV. However, as was noted in Ref.~\cite{Bajc:1996kj,Bajc:1996id}, the $\phi_1$ mass also receives a positive two-loop thermal correction from the third operator in Eq.~(\ref{eq:vphiR}). The same-order effect also comes from a two-loop correction with fermions, see Figure~\ref{fig:2_loop}. 
Computing these corrections one gets~\cite{Bajc:1996kj} 
\be\label{eq:deltavphiR2}
\delta V_T'=  \frac {9}{64} \frac{T^2}{\Lambda^2} T^2 \phi_1^2.
\ee 
Requiring  that the positive correction of (\ref{eq:deltavphiR2}) is subleading compared to the negative mass correction from~(\ref{eq:deltavphiR1}), implies
\be\label{eq:phiRsnr2}
T^2 < \frac 8 3 \mu \Lambda.
\ee
Given that the upper bound on the temperature~(\ref{eq:phiRsnr2}) is very close to the lower one~(\ref{eq:phiRsnr1}), one should analyse them more carefully.
Adding all the mass corrections together we obtain the effective thermal mass
\be
m_{\phi_1}^2(T) = \mu^2 - \frac 3 4 \frac {\mu}{\Lambda} T^2 + \frac 9 {32} \frac{T^4}{\Lambda^2} = \frac 1 2 \mu^2 + 2 \left( \frac 1 2 \mu - \frac 3 8 \frac {T^2}{\Lambda} \right)^2 > 0,
\ee
hence no SNR is actually possible.

The two-loop fermionic effect growing with temperature is linked to the nonrenormalizable fermion-scalar interactions, which, by Eq.~(\ref{eq:summ12_2}) are the necessary requirement for SNR.
As for the $|\phi|^6$ term in scalar potential, which also induces the positive thermal two-loop mass correction, its presence is required by supersymmetry, and can also be related to the need for having a scalar potential bounded from below, since the $\phi_1$ quartic term is negative~\cite{Bajc:1996kj}. Moreover, a scalar potential generated from any superpotential satisfies the boundedness condition by construction. This latter argument led the authors of Ref.~\cite{Bajc:1996kj} to suggests that the problem is generic to all supersymmetric theories, although no robust proof of that was presented. 
 
 \subsection{Symmetry Non-Restoration from High Multiplicity}
 
We will now show how the upper and lower bounds on SNR temperature in nonrenormalizable theories can be made compatible, by assuming a large number of states generating the thermal corrections.  
Let us consider two sets of superfields: $H_{u,d}$, whose scalar component VEVs would break EW symmetry, and $\chi_{1,2}$  transforming in a fundamental and an antifundamental representations of a $U(n)$ group, which will be responsible for the generation of the negative thermal mass to the Higgs fields. We will first use a somewhat simplified model allowing to demonstrate the mechanism, and will turn to the realistic case in the next section. The superpotential reads
\be\label{eq:ngt1}
W= \mu_\chi (\chi_1.\chi_2) + \frac{c_{\chi h}}{\Lambda} (H_u.H_d)\, (\chi_1.\chi_2),
\ee
and corresponding scalar potential is
\bea\label{eq:vscalhchi-1}
V =& \mu_\chi^2 (|\chi_1|^2+|\chi_2|^2)  
+ \frac{1}{\Lambda} c_{\chi h} \mu_\chi (|\chi_1|^2+|\chi_2|^2) (H_u.H_d+{\rm h.c.}) \label{eq:vscalsimp-1}\\
&+ \frac{c_{\chi h}^2}{\Lambda^2} \left\{ (|H_u|^2+|H_d|^2) |\chi_1.\chi_2|^2 + (|\chi_1|^2+|\chi_2|^2) |H_u.H_d|^2 \right\}.
 \label{eq:vscalsimp}
\eea
Repeating the steps performed in the previous example, we find that the $\chi$ fields introduce a  thermal correction to the $H_u.H_d$ mass mixing term, with a parametric size $n_\chi c_{\chi h} (\mu_\chi/\Lambda) T^2$. The appearance of the factor of $n_\chi$ can be simply understood from Eq.~(\ref{eq:vscalsimp-1}), where to evaluate the thermal average one needs to sum over all the $\chi_{1,2}$ components, $\langle |\chi_{1,2}|^2 \rangle_T \sim n_\chi T^2$. 
This mixing results in one of the Higgs mass eigenstates developing a negative mass and rolling away from the origin, thus breaking the EW symmetry.  
For this to happen, the Higgs mass mixing $\sim n_\chi c_{\chi h} (\mu_\chi/\Lambda) T^2$ has to overcome the positive thermal correction to the diagonal Higgs mass $\delta m_h^2(\text{SM})\sim {\cal O}(1) T^2$ induced by the SM states, implying
\be\label{eq:snr1}
\frac{c_{\chi h} \mu_\chi}{\Lambda} \gtrsim \frac 1 {n_\chi}.
\ee
On the other hand, the dimension-six terms in Eq.~(\ref{eq:vscalsimp}) generate positive two-loop thermal contributions to the diagonal Higgs mass matrix elements $ \sim n_\chi c_{\chi h}^2 T^4/\Lambda^2$, which force the  EW symmetry restoration at very high temperature. Analogous two-loop contribution comes from the loops of fermionic $\chi$ components, as depicted in Figure~\ref{fig:2_loop}. These two-loop corrections are subdominant compared to the off-diagonal Higgs thermal mass term as long as 
\be\label{eq:snr2}
\frac{c_{\chi h} \mu_\chi}{\Lambda} \lesssim \frac{\mu_\chi^2}{T^2}.
\ee
Combination of the constraints~(\ref{eq:snr1}) and~(\ref{eq:snr2}) provides an upper bound on the temperature, $T\lesssim \mu_\chi \sqrt {n_\chi}$.
At the same time, for the thermal effects induced by the SNR states to be efficient, their mass should not be much greater than the temperature, $\mu_\chi^2 \lesssim T^2$, hence SNR can occur in the interval of temperatures
\be\label{eq:snr3}
|\mu_\chi| \lesssim T \lesssim \sqrt{n_\chi} |\mu_\chi|.
\ee
Thus while for $n_\chi=1$ the `no go' theorem of \cite{Haber:1982nb,Mangano:1984dq,Bajc:1996kj,Bajc:1996id} hold, the conclusion that SNR is impossible is avoided in the case that $n_\chi\gg1$, allowing for a modest window in which to realize SNR.
Notably, this feature is analogous to what happens in the model of  fermions-induced SNR~\cite{Matsedonskyi:2020mlz,Matsedonskyi:2020kuy,Matsedonskyi:2021hti}, which is included in the supersymmetric model considered here as a sub-sector. 

Finally, we note that while a generic non-renormalizeable theory is expected to feature an infinite set of higher-dimensional operators, their effect can be kept under control in the large-$n_\chi$ limit. We discuss this point, together with higher-order thermal loops, in Appendix~\ref{sec:higherloopphi12}.

\section{Electroweak Symmetry Non-Restoration with Supersymmetry}\label{sec:ewsnr}

We will now discuss the SNR mechanism presented in the previous section in more detail. Following the usual approach, we will assume that the $\chi$ fields coupled to the Higgs sector are new SM-singlets. Introducing new states for the purpose of SNR (rather than trying to use superpartners of the SM fields) allows us to choose freely their couplings and multiplicity. Notably, the new particles, being SM singlets, can have a mass near the EW scale without conflict with experimental data (unlike, for instance, squarks). This lightness allows them to effectively contribute to the Higgs thermal potential around $T\simeq 100$~GeV, where the SM thermal effects would otherwise shift the Higgs VEV below the critical value $h/T=1$. As for the $U(n_\chi)$ symmetry, for simplicity we will restrict our analysis to the case when it remains unbroken at all relevant temperatures.

Our main interest here is to demonstrate the possibility of EW SNR, hence we will pay maximal attention to the SNR sector and its couplings to the Higgs. The Higgs sector itself is, however, beyond our main focus.  We will make the simplifying assumption that the Higgs sector is in the alignment limit~\cite{Gunion:2002zf} (as we detail shortly), and also take the masses of the additional Higgs states to be $\sim2$~TeV~\cite{Atkinson:2021eox}, which is beneficial for reproducing experimental observations, but without specifying how the required masses and couplings are generated. We will also neglect the thermal effects of the SM fields' superpartners.
We will consider the following superpotential
\be\label{eq:ngt2}
W= \mu_\chi (\chi_1.\chi_2) + \mu_h (H_u.H_d) + \frac{c_{\chi h}}{\Lambda} (H_u.H_d)\, (\chi_1.\chi_2) + \frac{c_{\chi}}{\Lambda} (\chi_1.\chi_2)^2,
\ee
where $H_u$ and $H_d$ are the two MSSM Higgs superfields, and $\chi_1, \chi_2$ are new SM singlet chiral superfields transforming as a fundamental and an antifundamental representations of a $U(n_\chi)$ symmetry. $H_u.H_d$ is a shorthand notation for the $SU(2)_L$-invariant term $(H_u)^\alpha \epsilon^{\alpha \beta} (H_d)^\beta$. 
Additionally, we will assume the following overall Higgs potential
\bea\label{eq:v2hdm}
V_{\rm 2HDM} &=& m_{Hu}^2|H_u|^2 + m_{Hd}^{2}|H_d|^2+m_{Hud}^2(H_u.H_d + {\rm h.c.})\\
&+&\frac{\beta_1}{2}|H_d|^4+\frac{\beta_2}{2}|H_u|^4+\beta_3 |H_u|^2 |H_d|^2 + \beta_4 |H_u.H_d|^2, 
\eea
where the parameters $m_i^2,\beta_i$ can be traded for the physical parameters:
\be
m_{h,H,H_+,H_A}, \qquad \tan \beta = \frac{v_u}{v_d}, \qquad  v^2=v_u^2+v_d^2,
\ee
 and the mixing angle $\alpha$~\cite{Bhattacharyya:2015nca}, where $\langle H_u \rangle = (0,v_u/\sqrt 2)$, $\langle H_d \rangle = (v_d/\sqrt 2,0)$ such that $v_u=v\sin\beta$ and $v_d=v\cos\beta$.

The scalar potential resulting from Eq.~(\ref{eq:ngt2}) is
\bea\label{eq:vscalhchi}
V &=& \mu_\chi^2 |\chi_i|^2 + \mu_h^2 |H_u|^2 + \mu_h^2 |H_d|^2 \nn\\
&+& \frac{1}{\Lambda} \left\{ c_{\chi h} \mu_\chi |\chi_i|^2 (H_u.H_d+{\rm h.c.}) 
+c_{\chi h} \mu_h |H_i|^2 (\chi_1.\chi_2+{\rm h.c.}) + 2 c_\chi \mu_\chi |\chi_i|^2 (\chi_1.\chi_2+{\rm h.c.}) \right\} \nn\\
&+& \frac{c_{\chi h}^2}{\Lambda^2} \left\{ |H_i|^2 |\chi_1.\chi_2|^2 + |\chi_i|^2 |H_u.H_d|^2 \right\}
+\frac{2 c_{\chi h} c_{\chi}}{\Lambda^2} \left\{ |\chi_i|^2 (H_u.H_d \chi_1^\dagger.\chi_2^\dagger + {\rm h.c.}) \right\} \nn\\
&+&\frac{4 c_{\chi}^2}{\Lambda^2} \left\{ |\chi_i|^2 |\chi_1.\chi_2|^2 \right\},\label{eq:vscal6}
\eea
where $i=1,2$ for $\chi$ and $i=u,d$ for $H$ fields.
The fermionic part of the Lagrangian reads
\begin{equation}
\label{eq:lagferm}
-{\cal{L}}_{F} = \mu_\chi \, \tilde \chi_1.\tilde \chi_2 
+ \frac{1}{\Lambda} \left(c_{\chi h} \tilde \chi_1.\tilde \chi_2 \, H_u.H_d  
+ c_\chi(\tilde \chi_1.\chi_2+\tilde \chi_2.\chi_1)^2 + 2c_\chi \, \chi_1.\chi_2 \, \tilde \chi_1.\tilde \chi_2  + {\rm h.c.}\right).
\end{equation}
Notably, the term $c_{\chi h} \tilde \chi_1 \tilde \chi_2 H_u.H_d/\Lambda$ gives a thermal contribution to the Higgs mass mixing with magnitude $\sim c_{\chi h} \mu_\chi T^2 /\Lambda$, furthermore, a two-loop diagram with the same interaction generates a correction to the diagonal mass $\sim c_{\chi h}^2 T^4/\Lambda^2$. 

In the following sections we will discuss more quantitatively the effect of the new states on the Higgs thermal mass, and the conditions needed to achieve EW SNR.

\subsection {Higgs and $\chi$ thermal potential}

The one-loop thermal correction to the scalar potential can be computed using the standard expression involving sums over bosons $b$ and fermions $f$
\be\label{eq:1loopvh}
\delta V_T =  \sum_b g_b \frac{T^4}{2\pi^2} J_b[{m_b^2}/{T^2}] - \sum_f g_f\frac{T^4}{2\pi^2} J_f[{m_f^2}/{T^2}],
\ee 
where $g_b$ and $g_f$ are numbers of bosonic and fermionic degrees of freedom, $m_{b,f}^2$ are their masses, and the thermal loop functions are
\be
J_b[x]= \int_0^\infty dk\, k^2 \log \left[1- e^{-\sqrt{k^2+x}} \right],\qquad
J_f[x]= \int_0^\infty dk\, k^2 \log \left[1+ e^{-\sqrt{k^2+x}} \right].
\ee
While the exact expression will be used for deriving the numerical results, we will first use the high-$T$ approximation, $m_{b,f}^2/T^2\ll1$, to understand the analytic behaviour of the thermal effects. In this case the thermal corrections are simplified to the following expression 
\be
\delta V_T \simeq \frac {T^2}{24} \left[\sum_b g_b m_b^2 + \frac 1 2 \sum_b g_f m_f^2 \right].
\ee 
Note that this expression was used earlier in Eq.~(\ref{eq:deltaVT1}).

Using the scalar potential~(\ref{eq:vscalhchi}), the fermionic part of the Lagrangian~(\ref{eq:lagferm}) and the standard form of the interactions of $H_{u,d}$ with the top quark and SM gauge bosons, we obtain the following expression for the sum of the fermion mass terms, retaining only the terms quadratic in fields 
\begin{equation}
\sum_{f} g_f m_f^2 = 12 {y_t}^2|H_u|^2 +
4 n_\chi \frac{c_{\chi h} \mu_\chi}{\Lambda} (H_u.H_d+ {\rm h.c.}) + 
 8 \frac {c_{\chi} \mu_\chi (n_\chi+1)}{\Lambda} (\chi_1.\chi_2+{\rm h.c.}),
\label{eq:summf} 
\end{equation}
and for the boson mass terms
\bea
\sum_{b} g_b m_b^2 &=& 2 (3\beta_2 + 2 \beta_3 + \beta_4) |H_u|^2 + 2 (3\beta_1 + 2 \beta_3 + \beta_4)  |H_d|^2 + \frac 3 2 (3 g^2 + g^{\prime 2})(|H_u|^2+|H_d|^2) \nonumber\\
&+& 4 n_\chi \frac{ c_{\chi h}  \mu_\chi}{\Lambda} (H_u.H_d+{\rm h.c.}) 
+ 8\frac{c_\chi \mu_\chi (n_\chi+1)+c_{\chi h} \mu_h}{\Lambda} (\chi_1.\chi_2+{\rm h.c.}),
\label{eq:summb}
\eea 
where $y_t$ is the top quark Yukawa and $g,g'$ are the EW gauge couplings.
These thermal corrections can be accounted for by modifying the mass terms in the Higgs potential~(\ref{eq:v2hdm}) in the following manners
\bea
m_{Hu}^2 &\to& m_{Hu}^2 +  \frac{y_t^2}{4} T^2  + \frac{T^2}{12} (3\beta_2 + 2 \beta_3 + \beta_4) +  \left( \frac {3 g^2}{16} + \frac { g^{\prime 2}}{16} \right) T^2 \equiv m_{Hu}^2+c_{Tu} T^2  \nn\\
m_{Hd}^2 &\to& m_{Hd}^2 +  \frac{T^2}{12} (3\beta_1 + 2 \beta_3 + \beta_4) +  \left( \frac {3 g^2}{16} + \frac { g^{\prime 2}}{16} \right) T^2 \equiv m_{Hd}^2+c_{Td} T^2 \nn\\
m_{Hud}^2 &\to& m_{Hud}^2 +  n_\chi \frac{c_{\chi h} \mu_\chi}{4\Lambda} T^2 ,
\label{eq:mheff}
\eea
while the thermally corrected $\chi_i$ mass eigenvalues read
\be\label{eq:mchiT}
m_{\chi \pm}^2(H_i)_T = \mu_\chi^2 + \frac{c_{\chi h} \mu_\chi}{\Lambda} (H_u.H_d+{\rm h.c.})
\pm  \left\{ \frac{c_{\chi h} \mu_h}{\Lambda} (|H_u|^2+|H_d|^2) + \frac{c_\chi \mu_\chi (n_\chi+1)}{2\Lambda} T^2  + \frac{c_{\chi h}\mu_h}{3\Lambda} T^2 \right\}.
\ee
We will now use these expressions to analyse the thermal evolution of our model.

\subsection {Electroweak Symmetry Non-Restoration}
\label{sec:3.2}

Let us now derive more precisely the conditions needed to obtain EW SNR. To this end we will examine the mass matrix of the Higgs doublet components $h_u,h_d$
\be
{\cal M}^2_h = 
\left[\begin{matrix}
m_{Hu}^2 (T) & m_{Hud}^2(T)  \\
m_{Hud}^2(T) & m_{Hd}^2 (T)
\end{matrix}\right],
\ee
where the mass matrix elements are defined in Eq.~(\ref{eq:mheff}). High-$T$ EWSB can be achieved when one of the mass eigenvalues becomes negative, hence we need the determinant of the mass matrix to be negative as well, implying
\be
m_{Hud}^4(T)>m_{Hu}^2 (T)  m_{Hd}^2 (T),
\ee 
or, explicitly,
\be
\left|m_{Hud}^2 + n_\chi \frac{c_{\chi h} \mu_\chi}{4\Lambda} T^2 \right| > \sqrt{(m_{Hu}^2+c_{Tu} T^2)( m_{Hd}^2+c_{Td} T^2)}.
\ee
This condition then places a lower bound on the combination $n_\chi c_{\chi h} \mu_\chi/\Lambda$. 
Let us consider this expression in the limits of high and low temperature.
At very high $T$, where all the $T$-independent contributions are negligible, the SNR condition becomes
\be
\left|n_\chi \frac{c_{\chi h} \mu_\chi}{4\Lambda} \right| > \sqrt{c_{Tu} c_{Td}}.
\ee

\begin{figure}[t]
\includegraphics[width=8.cm]{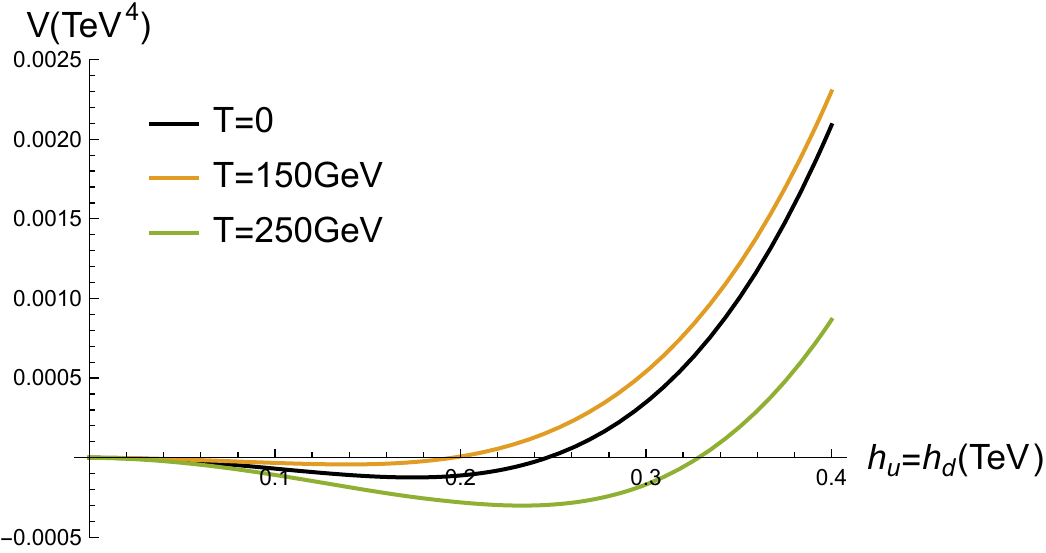}
\hspace{0.5cm}
\includegraphics[width=7.cm]{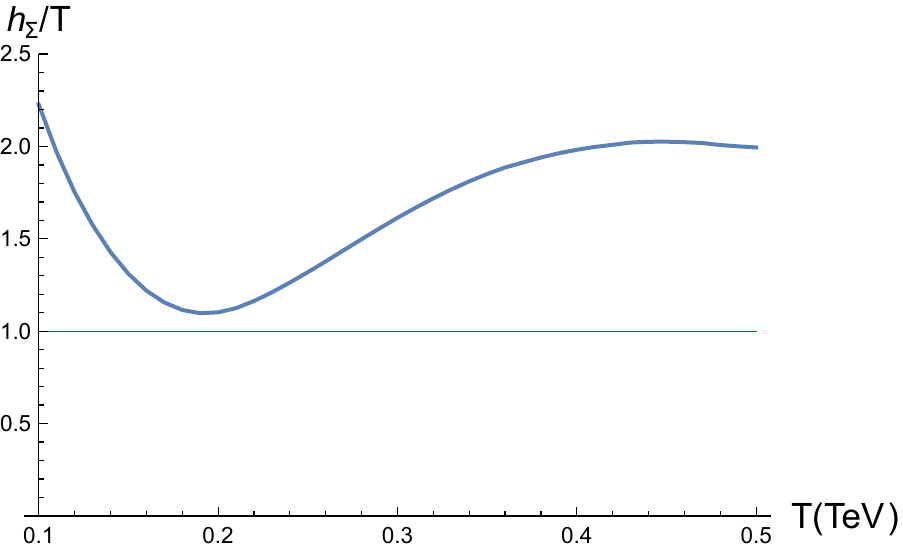}
\caption{{\it Left panel:} The Higgs potential in the $h_u=h_d$ direction at various temperatures in the presence of SNR states, for $n_\chi=10, \mu_\chi=650~\text{GeV}, \mu_H=-130~\text{GeV}, \Lambda=2.5~\text{TeV},c_{\chi h} = 3.2, \tan \beta =1$ and a common heavy Higgs mass scale $m_{H,H_+,H_A}=2~\text{TeV}$. {\it Right panel:} The corresponding Higgs VEV $h_\Sigma=\sqrt{h_u^2+h_d^2}$ evolution with temperature.}
\label{fig:vhexamples}
\end{figure}

Experimental data \cite{Haber:2013mia} suggests that the additional Higgs states are heavy, and the neutral Higgs mixing is close to the alignment limit \cite{Gunion:2002zf}. In this case, to study EWSB at low temperatures we can just rotate the states to the mass eigenbasis and consider the VEV of the lightest state $h_{\rm light}$.  The thermal correction to the potential of this light state induced by the  SNR fields is given by
\bea\label{eq:snralign}
V_{\rm SNR} \,=\, H_u.H_d \frac{n_\chi c_{\chi h} \mu_\chi}{\Lambda} \frac{T^2}{4}  + {\rm h.c.} 
\,\to\,  - c_\alpha s_\alpha h_{\rm light}^2 \frac{n_\chi c_{\chi h} \mu_\chi}{\Lambda}  \frac{T^2}{4},
\eea
where in the second step we performed the rotations
\bea
h_u \to  c_\alpha h_{\rm light} - s_\alpha h_{\rm heavy}, 
\qquad h_d \to  s_\alpha h_{\rm light} + c_\alpha h_{\rm heavy}~,
\eea
 to the mass eigenstates basis.
Equation~(\ref{eq:snralign}) shows that the negative thermal mass of $h_{\rm light}$ is maximal for $c_\alpha  s_\alpha = 1/ 2$, which in the alignment limit $\cos (\alpha+\beta)=0$ implies $\tan \beta = 1$. Thus deviations from $\tan \beta = 1$ typically imply a reduction in the degree of SNR which can be achieved.
Since in the alignment limit the lightest mass eigenstate is SM-like, the positive thermal correction induced by the SM states is fixed to be
$\delta V_T \simeq 0.2\, T^2 h_{\rm light}^2.$
SNR can be achieved provided that these positive thermal corrections from SM states are 
overcome by the negative contribution of the SNR states (\ref{eq:snralign}), which implies the condition
\be\label{eq:snralign1}
s_{2\alpha} n_\chi   \frac{c_{\chi h} \mu_\chi}{\Lambda} \gtrsim 2.
\ee
In Figure~\ref{fig:vhexamples} we show an example of the resulting Higgs potential and the corresponding Higgs VEV evolution with temperature. 

\subsection {Regimes of Effective Field Theory Validity}\label{eft}

At high temperature the accuracy of our effective description can degrade for two reasons. The first one is the presence of higher-loop effects leading to thermal corrections which grow with temperature faster than the leading effect which we use for SNR. More precisely, we obtain (see Appendix~\ref{sec:twoloop})
\be
\delta V^{\rm 2-loop}_T \simeq 0.03\,   {n_\chi}  c_{\chi h}^2 \frac{T^4}{\Lambda^2} |H_i|^2.
\ee
Requiring these corrections to be small,  
$0.03\,   {n_\chi}  c_{\chi h}^2 T^4/\Lambda^2\ll T^2$,
 implies the bound
\bea\label{eq:eftbound1}
T&\ll& \frac{5}{\sqrt n_\chi} \frac{\Lambda}{c_{\chi h}}.
\eea
This condition also ensures the convergence of the series of the leading non-daisy thermal loops, see Appendix~\ref{sec:higherloopphi12}. 

On the other hand, our effective field theory (EFT) cannot capture the thermal effects induced by heavy physics at the scale $\Lambda$ which has been integrated out, hence for validity of the EFT one also needs to restrict the analysis to temperatures below the cutoff
$T\ll \Lambda$.
Numerically, the one-loop thermal effects of a state with mass $\Lambda$ become $\sim1/5$ 
suppressed for $T \lesssim \Lambda/3$. 
The restrictions of Eq.~(\ref{eq:eftbound1}) and $T\ll \Lambda$ are used in our numerical scans.

\subsection {Numerical scans}
\label{sec:scan}

In this section we will present the results of numerical parameter space scans of the model defined in Eq~(\ref{eq:ngt2}) to identify regions of EW SNR.
For now we will only be interested in finding the points which lead to EW symmetry being continuously broken from some high temperature down to $T=0$ with $h_\Sigma/T>1$, where $h_{\Sigma} = \sqrt{h_u^2 + h_d^2}$. The discussion of a possibility of having a high-temperature first-order EWPT is postponed to Section~\ref{sec:ewpt}.

In our numerical scans we used the exact one-loop thermal corrections~(\ref{eq:1loopvh}) improved with daisy resummation\footnote{We use all-mode daisy resummation~\cite{Parwani:1991gq} although resumming only the zero modes~\cite{Arnold:1992rz} does not change much our results. We use high-$T$ expansion for the computation of thermally corrected masses to be used for resummation, suppressing the contributions of heavy fields with a factor $\exp[-\nu m_i/T]$, with $\nu$ chosen to match numerically the one-loop thermal masses obtained without the high-$T$ expansion.}. In addition, to this we take into account zero-temperature 1-loop corrections via the Coleman-Weinberg potential \cite{Coleman:1973jx}. We impose the tree-level 2HDM stability bounds~\cite{Bhattacharyya:2015nca} on the scan points, and also check numerically that $h_i=v_i$ is the only minimum of the one-loop zero-temperature scalar potential within the field range $|h_i|,|\chi_i|<\Lambda$ (we could have allowed for additional metastable minima but we chose a stronger constraint to simplify the analysis). We require the thermal $\chi_i$ squared masses~(\ref{eq:mchiT}) to remain positive along the Higgs field trajectory,\footnote{This requirement is introduced to simplify the analysis since negative thermal squared masses for  $\chi_i$ would imply non-zero $\chi_i$ VEVs, greatly complicating the numerical calculation. This restriction could be dropped, potentially leading to alternative viable parameter points, but this is beyond the scope of this work. Weinberg's original work~\cite{Weinberg:1974hy} suggests it can be possible for the fields driving the high-$T$ symmetry breaking to have VEVs.} and that the two-loop and cutoff effects to be at least $1/5$ suppressed. We scan the Higgs VEVs' trajectories from $T=0$ and terminate at $T_{\text{max}}$ such that either $h/T$ drops below 1, or the two-loop and cutoff effects become too large, or one of the $\chi_i$ thermal squared masses becomes negative.

\begin{figure}[b!]
\includegraphics[width=3.67cm]{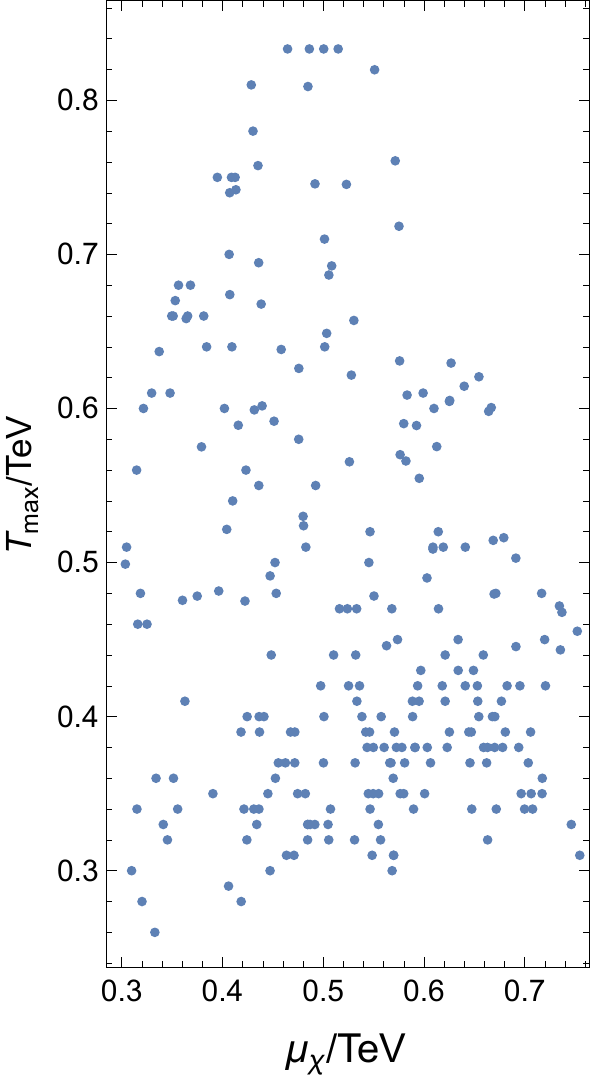}
\includegraphics[width=3.7cm]{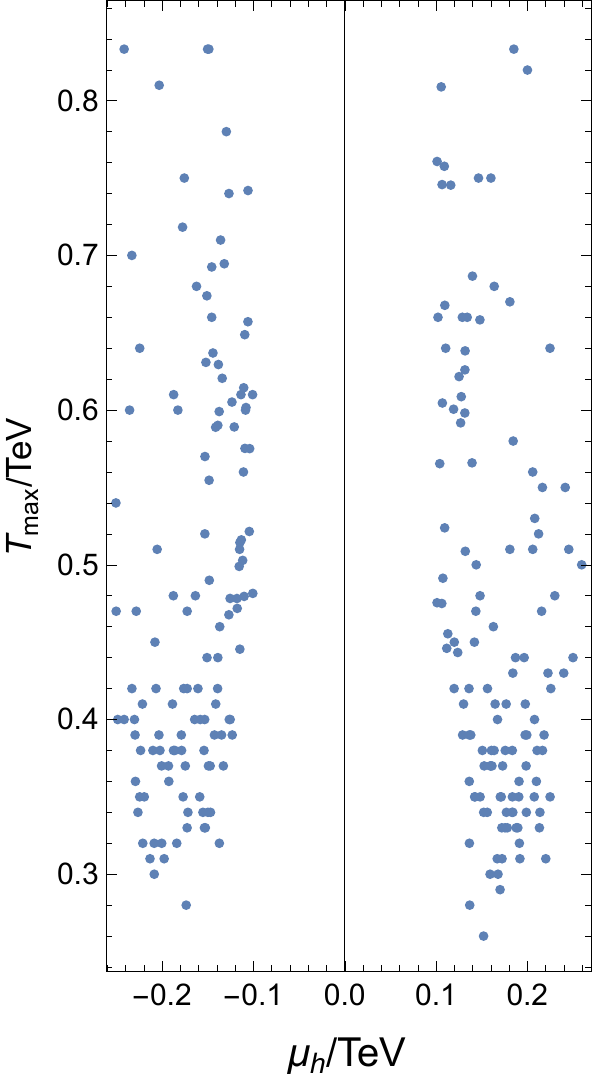}
\includegraphics[width=3.75cm]{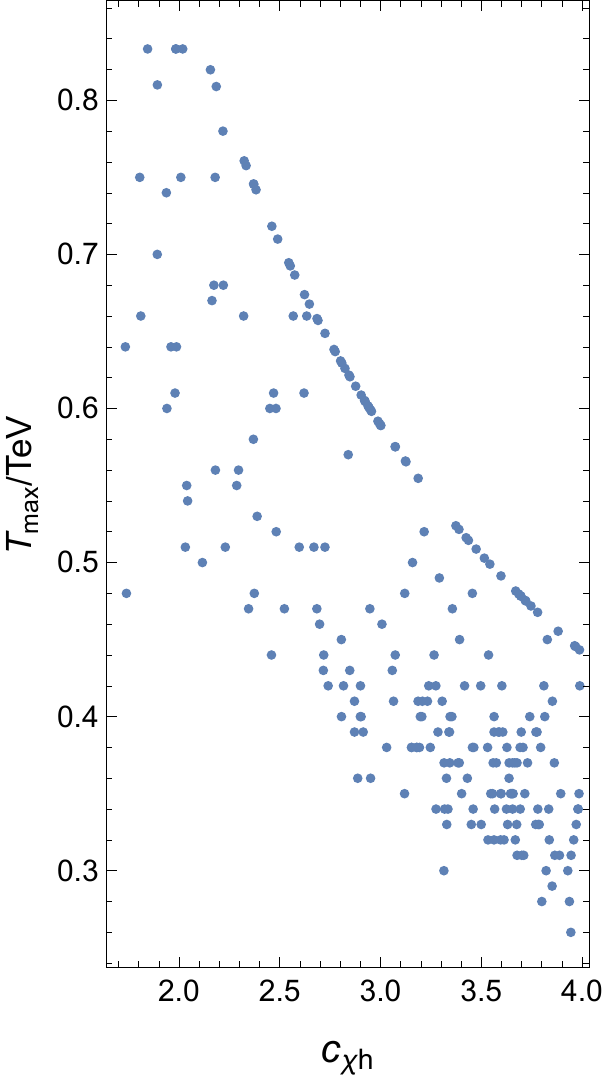}
\includegraphics[width=3.72cm]{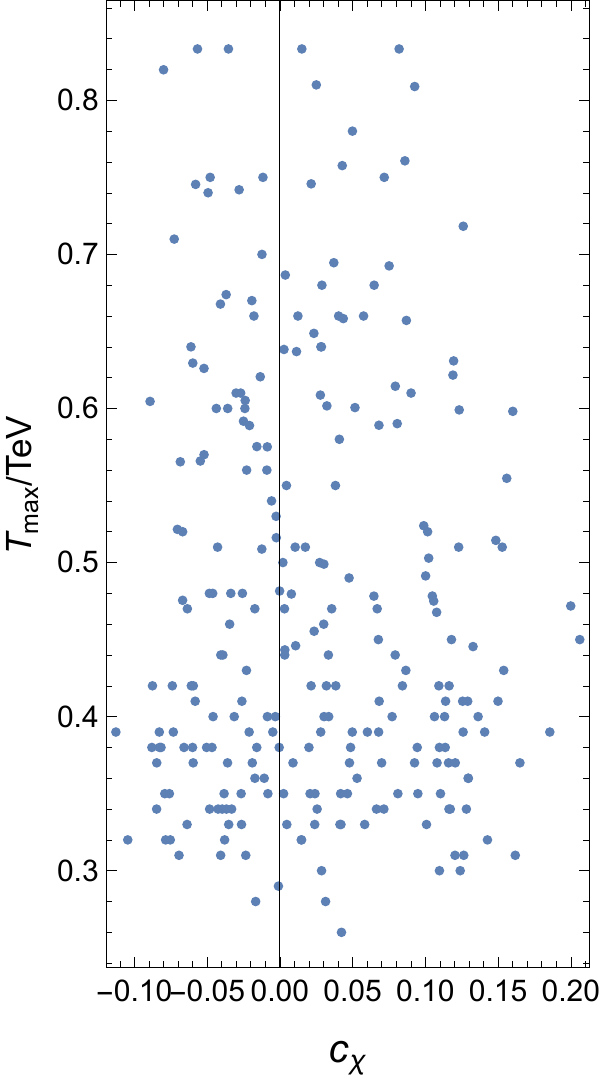}
\includegraphics[width=3.67cm]{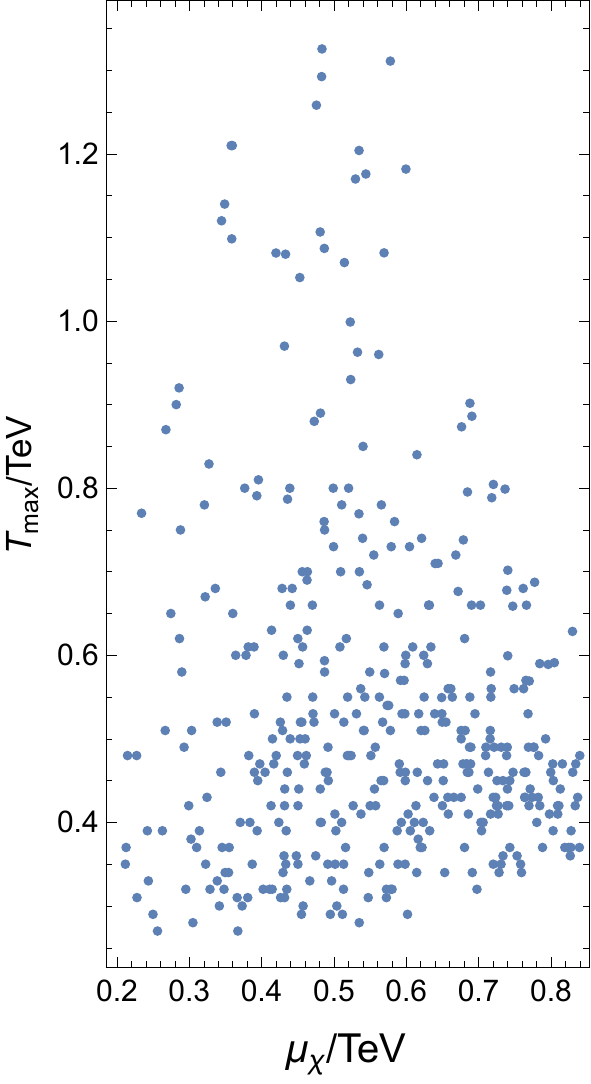}
\hspace{0.01cm}
\includegraphics[width=3.7cm]{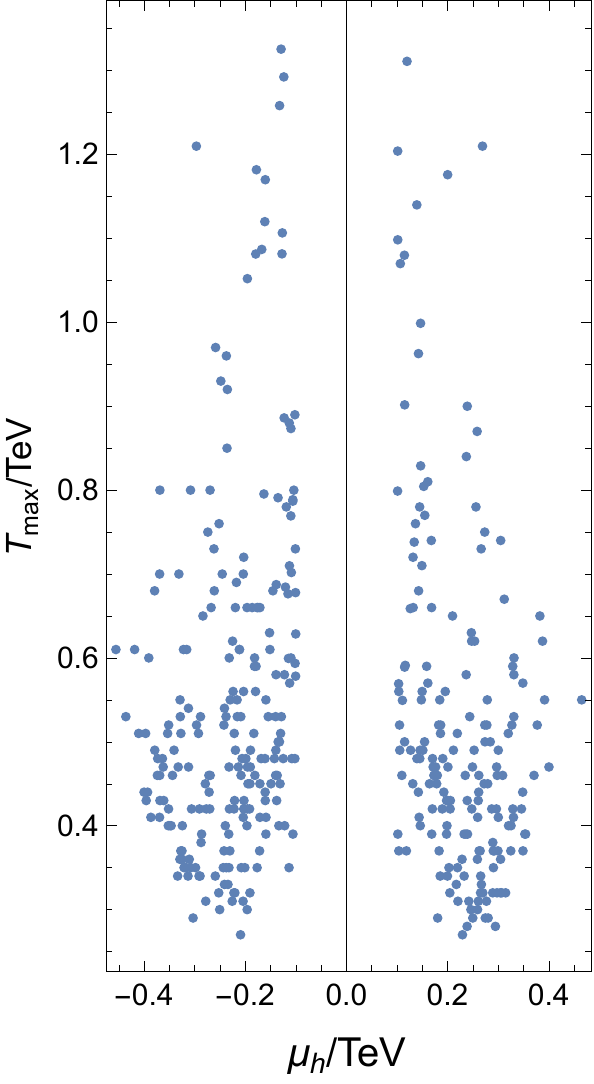}
\hspace{0.008cm}
\includegraphics[width=3.75cm]{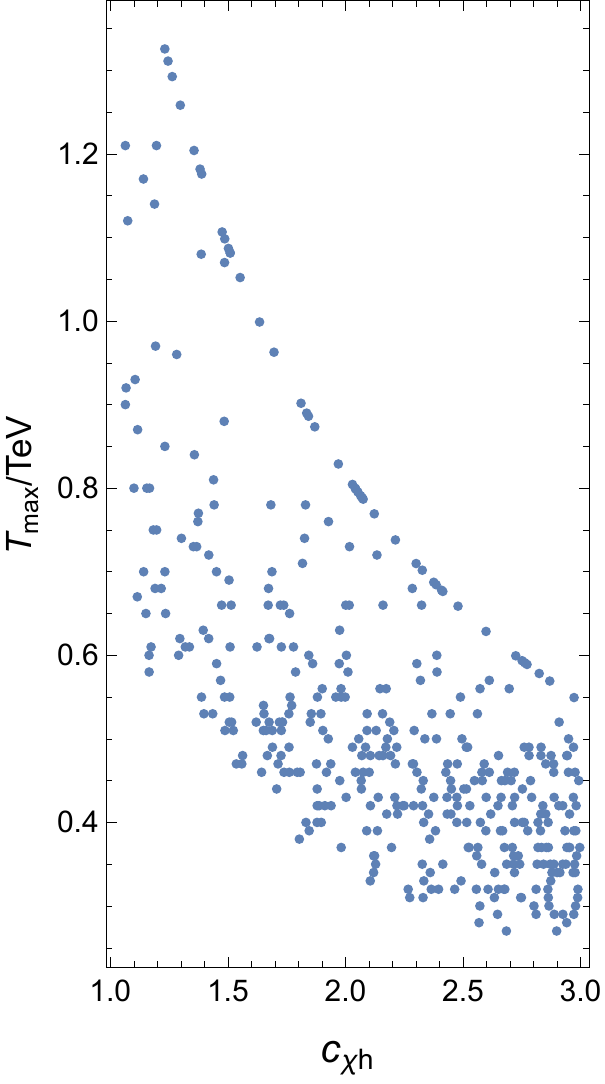}
\hspace{0.005cm}
\includegraphics[width=3.66cm]{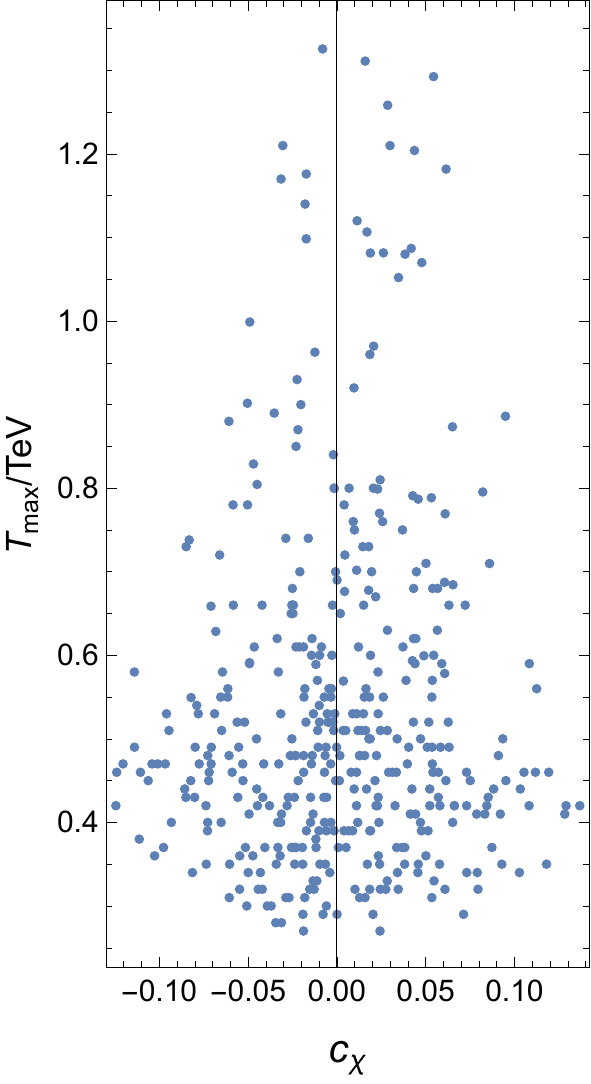}
\caption{Maximal SNR temperature as a function of $\mu_\chi$, $\mu_h$, $c_{\chi h}$ and $c_\chi$ where in the upper panels we fix $n_\chi=10, \Lambda=2.5~\text{TeV}, \tan \beta=1, m_{H,H_+,H_A}=2~\text{TeV}$ and in the lower panels we take $n_\chi=30, \Lambda=4~\text{TeV}, \tan \beta=1, m_{H,H_+,H_A}=2~\text{TeV}$.}
\label{fig:snrscansn20}
\end{figure}

The scan results are presented in Figure \ref{fig:snrscansn20}, where we show maximal SNR temperature as function of various parameters, for $n_\chi=10,30$. 
The maximal temperature first grows with $\mu_\chi$, since the latter controls the correction to the Higgs mass~(\ref{eq:snralign}), but then starts dropping since too large $\mu_\chi$ suppresses the density of $\chi$ particles in plasma.  The dependence on $c_{\chi h}$ is dictated by the perturbativity bounds~(\ref{eq:eftbound1}). The upper bounds on $c_\chi$ and $\mu_h$ are dictated by the requirement to have no additional minima of the scalar potential within the considered field range $|h_i|,|\chi_i|<\Lambda$, hence it can be relaxed by decreasing $\Lambda$, or by relaxing the constraint on additional minima. Note, the lower bound on $\mu_h$ is simply defined by our scan range.
In agreement with analytical estimates (cf.~Eq.~(\ref{eq:snr3})), the maximal temperature grows as $\sqrt{n_\chi}$. 
Notably, $n_\chi\sim10$ permits SNR such that EWSB is delayed until the TeV scale, presenting a new scale at which one might realize EW baryogensis.

\section{Generating the First-Order Electroweak Phase Transition}\label{sec:ewpt}

 Famously, Sakharov enumerated the conditions for successful baryogenesis\footnote{These criteria, while providing a general framework are not strictly necessary conditions, with Spontaneous Baryogenesis of Cohen \& Kaplan \cite{Cohen:1988kt} providing an example which does not conform to these requirements.} \cite{Sakharov:1967dj}: \begin{itemize}
\item[i)] Baryon number $B$ violation,  
\item[ii)] Violation of $C$ and $CP$ symmetries, 
\item[iii)] A period of out-of-equilibrium dynamics. 
\end{itemize}
EWBG is highly attractive since phase transitions can lead to a departure from equilibrium and EW sphalerons are a natural source of  $B$ violation.
 Notably, in the Standard Model the EW phase transition to the broken phase is via a smooth crossover, and thus is inadequate to satisfy Sakharov (iii). Moreover, while the CKM matrix presents CP-violating phases these are too small to account for the magnitude of the observed baryon asymmetry.

In Section~\ref{sec:ewsnr} we have demonstrated that the EW symmetry can stay broken starting from some TeV-scale temperature down to zero-temperature. However, to ensure appropriate out-of-equilibrium dynamics, successful models of EW baryogenesis require that the phase transition be strongly first-order, which now can potentially happen at $T \gg 130$~GeV. The criteria for the strength of the  first order   phase transition is typically taken to be \cite{Quiros:1999jp}
\be
\frac{h_\Sigma}{T}>1~,
\ee
for all relevant temperatures after the transition. Baryon asymmetries generated in models which do not satisfy this requirement will typically be aggressively washed-out via sphaleron processes~\cite{Patel:2011th}.

For the purpose of demonstration we will use a simple, although probably not the most minimal, manner to obtain a first-order EW phase transition at high temperature. To this end we add a set of symmetry-restoring (SR) superfields $\psi$ transforming under their own $U(n_\psi)$, and which produce a thermal correction to the Higgs potential with a minimum at $h_{\Sigma}/T<1$. This minimum will dominate at high temperatures, while at lower temperatures the Higgs fields will transit to the minimum generated by the SNR states $\chi$ with $h_{\Sigma}/T>1$.

To introduce this new `layer' of fermions we add the following terms to the superpotential 
\be
\delta W = \mu_\psi (\psi_1.\psi_2) + \frac {c_{\psi h}}{\Lambda} (\psi_1.\psi_2)(H_u.H_d),
\ee
such that the new fermions and scalars have masses
\be\label{eq:psimass}
|m_{\tilde \psi}|^2 = \frac{1}{2}(m_{\psi_+}^2+m_{\psi_-}^2) = \left|\mu_\psi + c_{\psi h} \frac{H_u.H_d}{\Lambda}\right|^2 \to \left|\mu_\psi - c_{\psi h} \frac{h_u h_d}{2\Lambda}\right|^2.
\ee 
The resulting thermal correction to the light Higgs mass reads
\bea\label{eq:sralign}
V_{\rm SR} \,=  - c_\alpha s_\alpha h_{\rm light}^2 \frac{n_\psi c_{\psi h} \mu_\psi}{\Lambda}  \frac{T^2}{4}.
\eea
In order for the new states to push the Higgs VEV towards the origin, one must choose $c_{\psi h} \mu_\psi/\Lambda$ to be negative. At the same time, to ensure that the $\psi$ fields do not affect the Higgs field evolution at temperatures below the phase transition, so not to disrupt the SNR mechanism, we will choose $m_\psi$ to be much greater than $m_\chi$.

\begin{figure}[t!]
\includegraphics[width=8.cm]{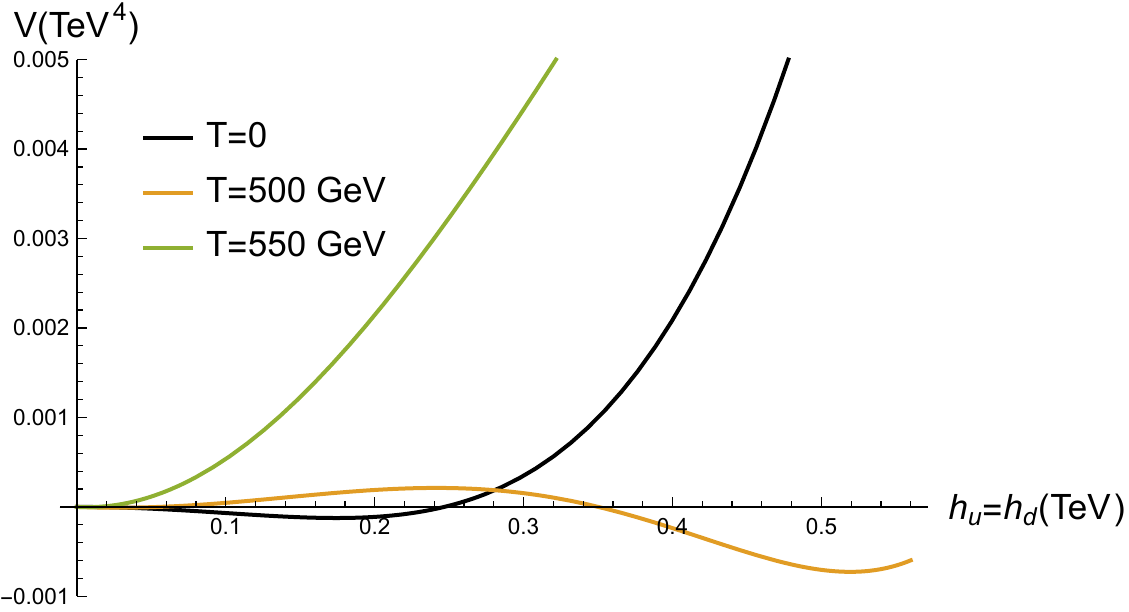}
\includegraphics[width=8.cm]{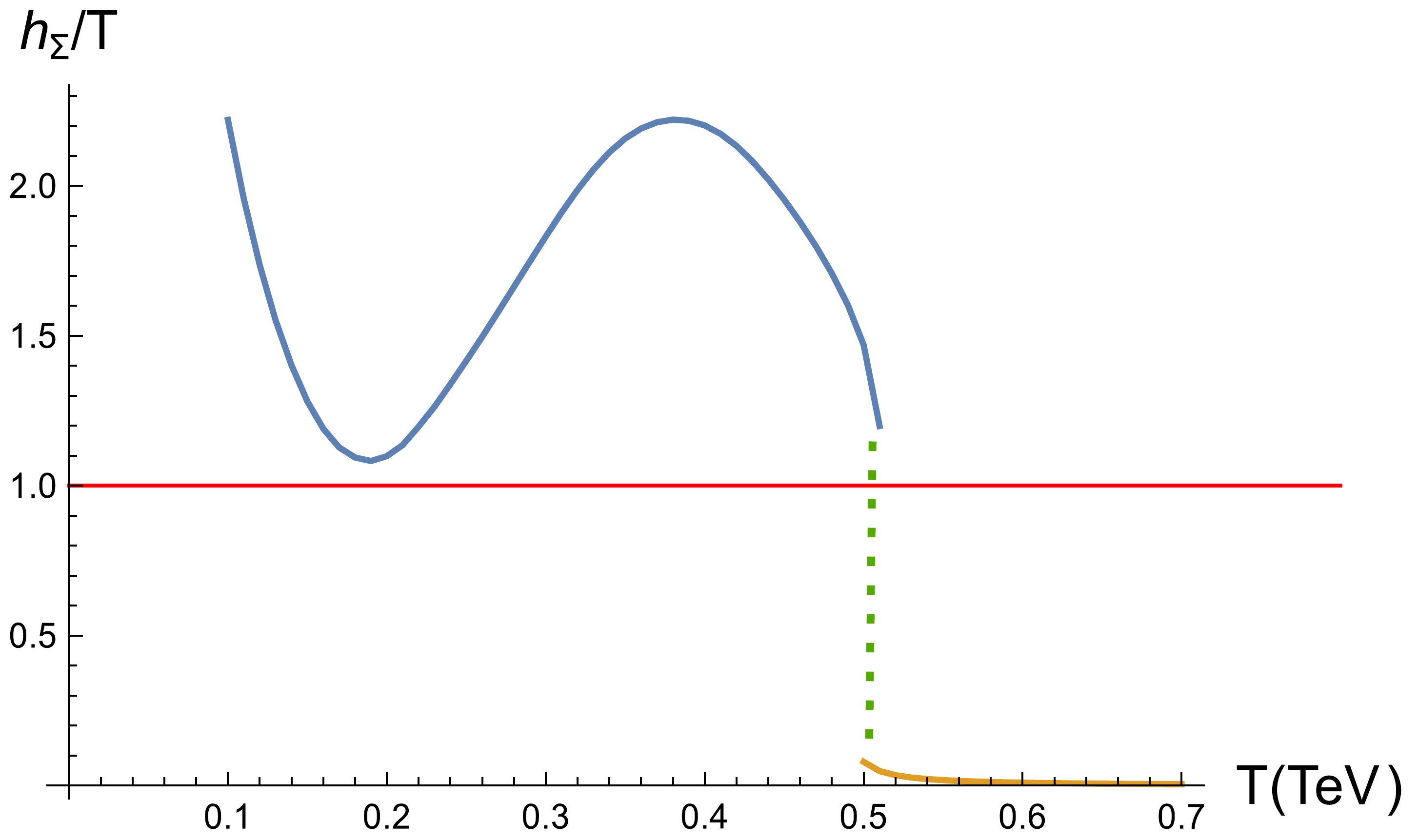}
\caption{{\it Left panel:} 
An example model in which the Higgs potential in the $h_u=h_d$ direction has a single minimum at zero temperature (black), two minima at $T=500$~GeV (orange), and a single minimum at $T=550$ GeV (green).
{\it Right panel:}  The evolution of $h_{\Sigma}(T)$ in this model with two coexisting minima and a phase transition at $T\simeq 500$ GeV marked by the dotted line.
For both panels we take  $n_\chi=n_\psi=10$, $\tan \beta =1$, $m_{H,H_+,H_A}=2~\text{ TeV},~\mu_\chi=0.77\text{ TeV}$, $\mu_\psi=1.5\text{ TeV}$, $\mu_H=150\text{ GeV},\Lambda=2\text{ TeV}$, $c_{\chi h}=3.7,~c_{\chi}=-0.2,~c_{\psi h}=-4.7$. We take the alignment limit in the Higgs sector and the MSSM superpartners to be above the scale $\Lambda$.}
\label{fig:ewpt2}
\end{figure}

\begin{figure}
\center
\includegraphics[width=8.25cm]{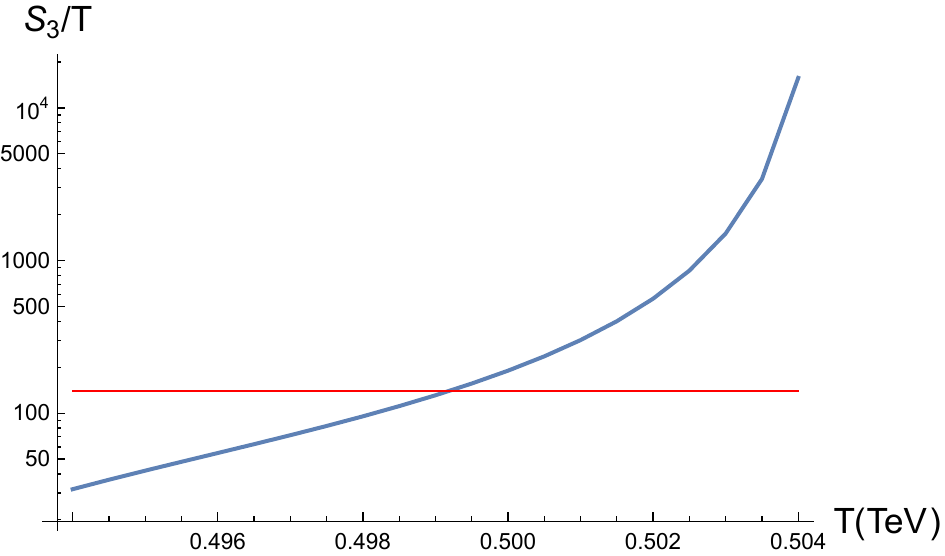}
\vspace{-1mm}
\caption{Bounce action $S_3/T$ dependence on the temperature (blue line) computed using the package FindBounce~\cite{Guada:2020xnz}. The tunneling rate starts dominating over expansion at $S_3/T\simeq 140$ (red line).
\vspace{-2mm}
}
\label{fig:S3}
\end{figure}

Figure~\ref{fig:ewpt2} (left) presents one such example with the Higgs potential featuring a transition between minima induced by $\psi$ and $\chi$ fields.  Figure~\ref{fig:ewpt2} (right) shows the corresponding thermal evolution of $h_\Sigma/T$ with a strongly first-order phase transition; the horizontal line indicates $h_{\Sigma}/T=1$; since the phase transition proceeds above this line it is considered strongly first order, and hence is suitable for EWBG.

A critical feature of a first order phase transition is that it occurs via bubble nucleation. As the thermal bath temperature falls below the critical temperature of EWPT, the bubble nucleation rate  $\Gamma_n(T)$ becomes non-zero and begins growing.  The phase transition to the broken phase occurs at temperature $T_n$ such that $\Gamma_n(T_n)$ overcomes the Hubble rate $H(T_n)$; the typical expectation for this to happen is
\bea
\left.\frac{S_3}{T}\right|_{T=T_n}\sim140,
\eea
where $S_3$ is the O(3) symmetric bounce action defining the bubble nucleation rate \cite{Linde:1981zj,Konstandin:2006nd}.
We numerically compute the bounce action $S_3/T$ dependence on the temperature using the Mathematica package FindBounce~\cite{Guada:2020xnz} for the model of Figure \ref{fig:ewpt2}. 
The temperature evolution of $S_3/T$ for this model is shown in Figure \ref{fig:S3}. Notably, this confirms that tunneling starts dominating over expansion at $S_3/T\simeq 140$, which occurs at approximately 500 GeV. Details the bubble nucleation can impact the efficiency of the asymmetry generation \cite{Balazs:2016yvi}.

 Figure \ref{fig:ewpt2} identifies one viable combination of parameters for a strongly first-order phase transition.  We next present some non-exhaustive numerical scans for the viable parameter space in which the EWPT is strongly first-order. Specifically, we scan around the successful example point for two different values of $\tan\beta=1$, 3. As discussed in Section \ref{sec:3.2}, deviations from $\tan\beta=1$ typically leads to a reduction in the negative Higgs mass correction, in order to compensate for this we increase $n_\chi$ and $n_\psi$ to 30 in order to maintain SNR for our scans with $\tan\beta=3$.

\begin{figure}[t!]
\includegraphics[width=7.5cm]{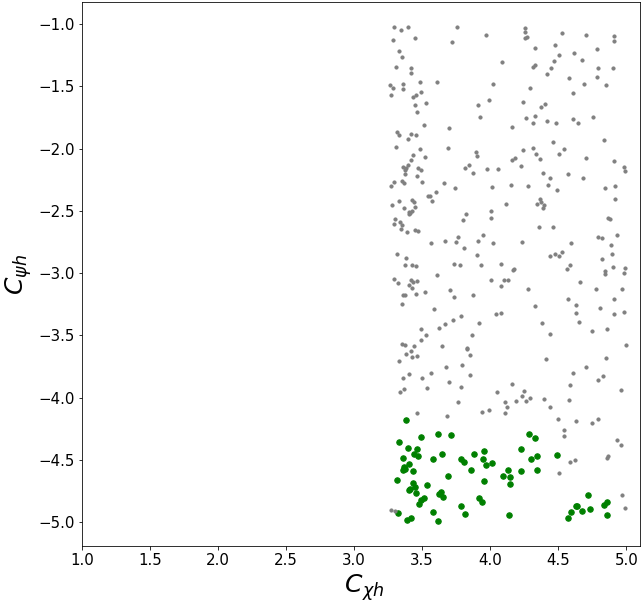}
\hspace{0.85cm}
\includegraphics[width=7.4cm]{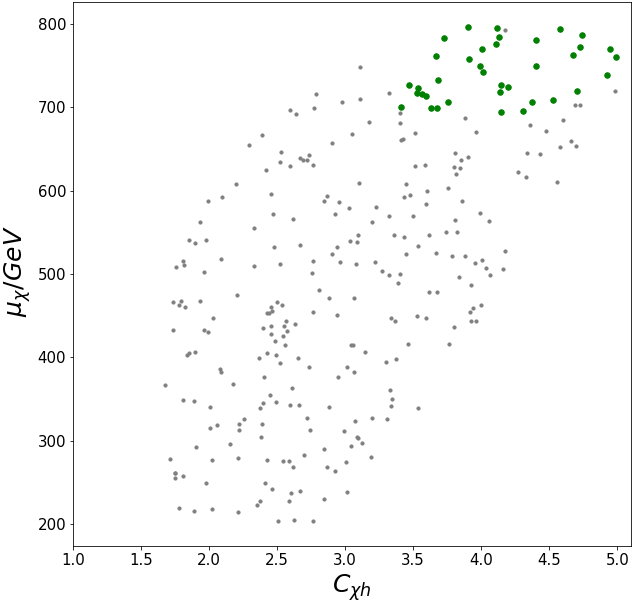}
\caption{Fixing $\tan \beta =1$, $n_\chi=n_\psi=10$. Left: We present non-exhaustive scans of the $c_{\chi h}$-$c_{\psi h}$ plane over the range $c_{\chi h}\in (1,5)$  and $c_{\psi h}\in(-1,-5)$, we fix the other parameters to be $m_{H,H_+,H_A}=2\text{ TeV},~\mu_\chi=0.77\text{ TeV}$, $\mu_\psi=1.5\text{ TeV}$, $\mu_H=150\text{ GeV},\Lambda=2\text{ TeV}$, $c_{\chi}=-0.2$. Right: We show scans of the $c_{\chi h}$-$\mu_{\psi}$ plane for $c_{\chi h}\in(1,5)$, and  $\mu_\chi/$GeV $\in(200,800)$ for the other parameters we take $m_{H,H_+,H_A}=2\text{ TeV}$, $\mu_\psi=1.5\text{ TeV}$, $\mu_H=150\text{ GeV},\Lambda=2\text{ TeV}$, $c_{\chi}=-0.2$, $c_{\psi h}=-4.7$. For both panels  green points indicate parameter values that pass all of our requirements, namely a strong-first order phase transition  at temperatures above 400 GeV. Grey points indicate parameter values for which SNR is achieved in excess of 200 GeV, but the phase transition is inappropriate for EWBG. \label{fig:SFOPTscan1}}
\vspace{10mm}
\includegraphics[width=7.5cm]{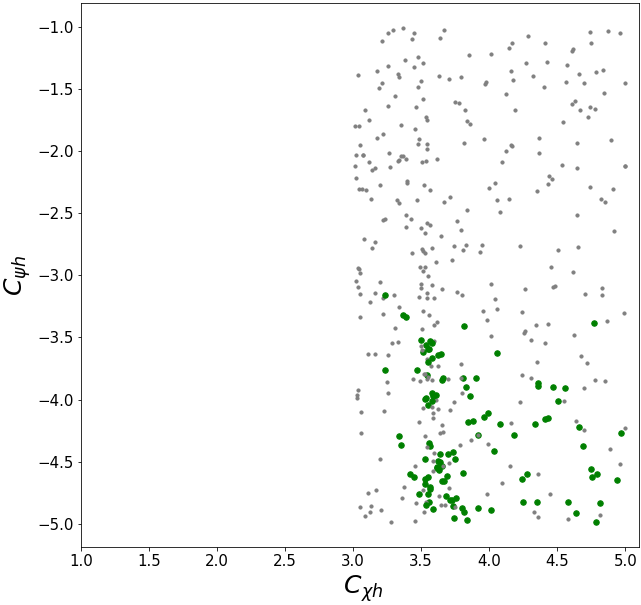}
\hspace{0.85cm}
\includegraphics[width=7.45cm]{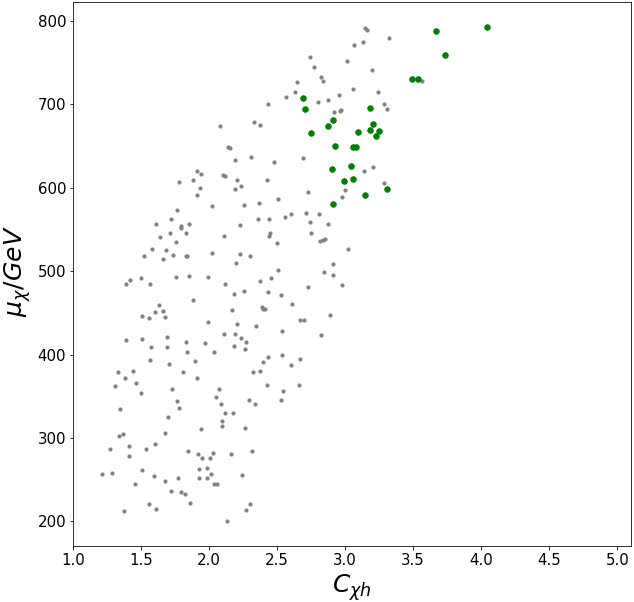}
\caption{As Figure \ref{fig:SFOPTscan1} but for $\tan \beta =3$ and $n_\chi=n_\psi=30$.}
\label{fig:SFOPTscan3}
\end{figure}

Starting with $\tan\beta=1$ and $n_\chi=n_\psi=10$ in Figure \ref{fig:SFOPTscan1} we present scans in the $c_{\chi h}$-$c_{\psi h}$ plane, and the $c_{\chi h}$-$\mu_{\psi}$ plane, keeping other parameter values the same as those in Figure \ref{fig:ewpt2}. We scan $c_{\chi h}$ over the range $(1,5)$, for $c_{\psi h}$ we scan over $(-1,-5)$, and for $\mu_\chi/$GeV over (200,800). Green points indicate parameter values that pass all of our requirements, namely a strong-first order phase transition occurring at temperatures above 400 GeV. Grey points indicate parameter values for which SNR is achieved in excess of 200 GeV, but the phase transition is not appropriate for EWBG. 
Let us now discuss qualitatively how the requirement of strong first order EWPT limits the available parameter space with respect to the space with SNR requirement only. Before starting it is important to mention that at relevant temperatures the $\psi$ states' mass is typically much larger  than the temperature and therefore the simple high-$T$ expansion can not be used to analyse the analytic properties of  the phase transition.   
The thermal barrier needed for the  phase transition is formed as a result of the thermal correction induced by the $\psi$ states $\delta V \propto \exp(-m_\psi(h)/T)$, which grows with $h$ (where $h$ is the direction in the $h_u-h_d$ plane along which the  phase transition occurs), and the falling correction induced by the $\chi$ fields $\delta V \propto - c_{\chi h} (\mu_\chi/\Lambda) h^2$. When the value of $|c_{\psi h}|$, controlling the Higgs-$\psi$ coupling decreases with respect to the benchmark value, so does the corresponding thermal correction, thus the thermal barrier disappears. In order to reintroduce the barrier, the temperature has to be raised, however this may lead to the temperature during the phase transition being larger than the value of the Higgs VEV in the broken minimum, hence the $h/T>1$ condition of strong first order phase transition is not satisfied. This results in the lower bound on $|c_{\psi h}|$ in the left panel of Figure~\ref{fig:SFOPTscan1}.
Furthermore, the decrease of $c_{\chi h}$ with respect to the benchmark value suppresses the negative contribution to the Higgs potential compared to the $\psi$-induced effect, which then makes the barrier too large, or turns the true minimum to a metastable one. To weaken the relative effect of the $\psi$ fields the temperature has to be lowered, however we require it to be at least $400$~GeV. As a result, $c_{\chi h}$ is limited from below as can be seen in the right panel of Figure~\ref{fig:SFOPTscan1}. Finally, if we decrease the value of $\mu_\chi$, 
the position of the minimum of the thermal potential induced by the $\chi$ states, $h_{\text{min}}^2 \propto \mu_\chi \Lambda/c_\chi$, decreases hence it becomes harder to satisfy the $h/T>1$ condition. This results in the lower bound on $\mu_\chi$ which can be observed in the right panel of Figure~\ref{fig:SFOPTscan1}.
Figure \ref{fig:SFOPTscan3}  shows analogous scans but taking $\tan\beta=3$ and $n_\chi=n_\psi=30$, with other parameters fixed as in Figure \ref{fig:ewpt2}.  The scans leading to Figures \ref{fig:SFOPTscan1}  \& \ref{fig:SFOPTscan3}  were performed using the public package CosmoTransition \cite{Wainwright:2011kj}.

\section{CP Violation in Supersymmetric Models}
\label{sec:cp}

EW sphalerons allow for $B$ violation, satisfying Sakharov (i), and in Section \ref{sec:ewpt} we identified settings in which a strong first order EWPT occurs at temperatures approaching $\sim1$ TeV, as required to satisfy Sakharov (iii). Thus, it remains to discuss potential sources of CP violation within the context of supersymmetric models as needed for Sakharov (ii). 

The premise of EWBG is that complex phases in parameters lead to CP-violating interactions of particles with the bubble walls formed by the first order phase transition, 
this results in a CP-asymmetric charge density in proximity of the wall. This CP-asymmetry diffuses ahead of the bubble wall and can be converted into other states through interactions in the plasma, leading to an excess density of left-handed antiparticles compared with their matter partners. 
The excess in left-handed states may then undergo sphaleron processes which are unsuppressed outside the bubble, such that baryon number is preferentially violated to create a net baryon density $\mu_{B_L}$ \cite{Cohen:1994ss}. Subsequently, this baryon asymmetry is transported inside to the interior of the expanding bubble, where it remains `safe' since sphaleron interactions are inactive in the broken phase. Once the phase transition completes the late stage baryon asymmetry $\eta$ becomes fixed. The source of the leading CP violation is an important detail in  determining the details of the mechanism. We will sketch some scenarios that we perceive as viable below, however,  a complete analysis of CP violating dynamics, leading to a computation of $\eta$, along with a comparisons to the experimental constraints (in particular, electric dipole moments \cite{Abel:2020pzs,ACME:2018yjb}) is beyond the scope of this work and requires a dedicated paper.

In identifying the source of CP violation it is important that it can be incorporated into our present setting without disrupting the earlier successes of SNR and the first-order nature of the EW phase transition. Moreover, one  must ensure that this CP violation is introduced in a fashion that satisfies experimental constraints. Notably, since in this setting the EWPT takes place at higher temperatures the states which introduce CP violation can be much heavier, which relaxes these experimental constraints.

A traditional method to introduce CP violation within the classic MSSM was via relative phases between the EW gaugino mass terms $M_{1,2}$ and the Higgsino $\mu_H$ term \cite{Carena:1996wj,Carena:2000id,Cline:1997vk,Cline:2000nw,Huet:1995sh}. These phases are of the form
$ \phi_i={\rm Arg}(\mu_H M_i b^*)$, with  ($i=1,2$),
where $b$ is the Higgs mass soft SUSY-breaking parameter.
Further, to evade constraints from EDM experiments one typically requires at least the first two sfermion generations to be very heavy, then the leading  constrains come from  two-loop diagrams involving charginos. In the MSSM this still leads to significant constraints on the parameter space \cite{Chang:2002ex,Li:2008kz}.

It was argued in \cite{Cirigliano:2009yd} that the ideal setting for supersymmetric EWBG in MSSM-like models required heavy sfermion masses $m_{\tilde{f}}\gtrsim 1$ TeV; relatively light pseudoscalar Higgs $A_0$, gauginos, and Higgsinos along with $\mu_H\sim M_i$ (for $i=1$ or $2$; this is called the `resonant neutralino/chargino baryogenesis funnel' \cite{Li:2008ez}); non-universal phases: $\phi_1\neq\phi_2$ and small to modest ${\rm tan} \beta$. Note that unlike pre-LHC MSSM models of EWBG we do not require a light ($\sim 100$ GeV) stop, since this is primarily introduced to arrange for a strong first order phase transition (which in our scenario is sourced via other fields). It should also be noted that for  ${\rm tan} \beta\gtrsim20$  the (s)bottoms and (s)taus alter the transport dynamics \cite{Cirigliano:2009yd,Chung:2009cb}, typically suppressing the final baryon asymmetry. Similarly, heavy pseudoscalar Higgs $A_0$ generically  lead to a suppression of the net $B$-number density generated at the EWPT \cite{Moreno:1998bq}. 

Moreover, the analysis of \cite{Cirigliano:2009yd} found that successful EWBG at the traditional critical temperature $T_c\sim100$ GeV could be obtained with $\mu_H\sim M_i$ (for $i=1$ xor  $i=2$), taking $M_2=2M_1\sim100$ GeV -- 700 GeV and $M_A\sim300$ GeV for relatively large (universal) CP phases $\phi_{1,2}\simeq0.3$. This paper assumed EWSB at $T_c\sim100$ GeV, while in our scenario we consider $T_c\sim500$ GeV, however as an initial proposal one might suppose to scale these masses together, as a naive guess of an appropriate spectrum. Since the EWPT occurs five-time higher in  Figure \ref{fig:ewpt2} (compared with traditional models), a similar scaling of the particle spectrum would imply $\mu_H\sim M_1\sim2.5$ TeV, $M_2\sim5$ TeV and $M_A\sim1.5$ TeV. Comparing to the spectrum of Figure \ref{fig:ewpt2}, the EFT cutoff and heavy Higgses were both taken to be $\Lambda=m_{H,H_+,H_A}=2$ TeV. Observe that the pseudoscalar Higgs masses are comparable, and since the Higgsinos and Gauginos (which source the CP violation) lie above the EFT cutoff $\Lambda$ this is unlikely to strongly disrupt earlier successes of SNR and the strong first order nature of EWPT. We note however that the $\mu_H$ in our EWPT scans is currently fixed to 150 GeV (cf.~Figure \ref{fig:SFOPTscan1}), and increasing the value much higher may require a corresponding increase in $n_\chi$ to maintain SNR. Thus, while encouraging, this certainly needs to be verified with explicit calculations of the final baryon asymmetry.

Given a specific model the baryon asymmetry can be calculated by evaluating the expression \cite{Cline:2000kb,Fromme:2006cm} 
\begin{equation}
\eta=\frac{405\Gamma_{\rm sph}}{4\pi^2v_wg_*T}\int_0^\infty{\rm d}z
\mu_{B_L}(z){\rm exp}\left(-\Gamma_{\rm sph}\frac{45z}{4v_w}\right)~,
\end{equation}
where $v_w$ is the wall velocity, $\Gamma_{\rm sph}\sim20 \alpha_W^5 T$ is the `weak' EW sphaleron rate, $g_*$ is the effective number of degrees of freedom. The exponential accounts for baryon number relaxation in case that the wall is slowly moving. From inspection of this form we note that the explicit $T_c$ suppression cancels against the temperature dependency in $\Gamma_{\rm sph}$. 

Finally, we note that the addition of the new singlet states for SNR/SR leads to an increase in $g_*$, it may also change the bubble wall velocity \cite{Moore:1995ua,Kozaczuk:2015owa,Azatov:2020ufh,Friedlander:2020tnq}, both of which impact $\eta$ and should be carefully checked in a full model. We anticipate that viable scenarios can be found for TeV-Scale Supersymmetric Electroweak Baryogenesis which reproduce the observed baryon asymmetry, however they likely still require modest to large CP violating angles, and as such a careful study comparing to EDM constraints will be necessary.


\section{Discussion}\label{sec:disc}

High-temperature symmetry breaking can have important consequences for a number of processes in the early universe. 
In particular, restoration or breaking of the electroweak symmetry above the electroweak scale can substantially affect the mechanisms of baryon asymmetry generation and their experimental tests. Here we have analysed the mechanisms of high-temperature symmetry breaking in supersymmetric theories and discussed a new way to overcome previously noted obstacles for symmetry breaking \cite{Haber:1982nb,Mangano:1984dq,Bajc:1996kj,Bajc:1996id}. We applied this mechanism to the supersymmetric extension of the Standard Model, showing that this can allow for a strong first order electroweak phase transitions, as required for electroweak baryogenesis, at scales significantly higher than the electroweak scale.
 
Raising the EWBG temperature to higher scales allows one to increase the mass scale of new physics involved in generating the asymmetry. However, the peculiar feature of our model is that the upper bound on the SNR temperature scales as the square root of the number of SNR states. As a result, assuming a moderate number of new states $n_\chi,n_\psi\sim10$, the EWBG temperature only increases by about one order of magnitude, weakening the potential signals of new physics at collider and CPV experiments and allowing one to evade the currently existing tensions, but not to the extent which would make the new physics completely undetectable at the foreseeable future experiments. 

There are several further improvements to our analysis that could be performed. We have assumed that all the superpartners of the SM states are too heavy to contribute to the Higgs thermal potential, which is not necessarily the case. In particular, there can be a situation when one of the stops is sufficiently light and contributes to SNR, taking part of the work done by the SNR sector.  Furthermore, our model can definitely be improved in the part related to the electroweak phase transition. In particular, it would be interesting to check if the SM superpartners or the heavy Higgs states can be used to generate the first order phase transition without invoking {\it ad hoc} symmetry-restoring states as we did.  

Our analysis concentrates on an effective field theory below some several-TeV scale $\Lambda$, hence it can be important to analyse possible UV completions to it.
In particular, for a renormalizable UV-completion we expect the EW symmetry to get restored at temperatures $T\gtrsim \Lambda$.~\footnote{The temperature at which the thermal effects of new physics at $\Lambda$ start playing a role and could drive the phase transition is $T=\Lambda/ \mathcal{O}(1)$, as discussed in the last paragraph of Section~\ref{eft}. This temperature can be lower than the temperature at which 2- and higher-loop effects of the EFT drive the symmetry restoration (Eq.~(\ref{eq:eftbound1})), if the parameter $c_{\chi h}$ is sufficiently low, while $n_\chi$ is not too large (which is something we are generally aiming at). Indeed, in our numerical scans we find that the there are regions of parameter space where the cutoff physics effects are expected to be relevant at temperatures which are lower than the temperatures at which higher-loop effects would be important.}
 If the transition from the restored to the broken phase can be arranged to be of the first-order, there might be no need for any additional physics to this end.  
 We also leave to future work the detailed analysis of the interplay between the baryon asymmetry and the bounds from the electric dipole moments.

\vspace{6mm}
\noindent {\bf Acknowledgments.}
The work of OM has been supported by STFC HEP Theory Consolidated grant ST/T000694/1. OM also thanks Mainz Institute for Theoretical Physics (MITP) and ICTP-SAIFR for their hospitality and support during completion of this work. JU is supported by NSF grant PHY-2209998 and wishes to thank the Berkeley Center for Theoretical Physics for their kind hospitality.

\appendix

\section{Higher-order thermal corrections} \label{sec:higherloopphi12}

To analyse higher-order thermal corrections to the Higgs and $\chi$ mass it is convenient to perform several simplifications of the Lagrangian. First of all, for simplicity we absorb $c_{\chi h}$ in $1/\Lambda$. Furthermore, we will set to zero the parameters $c_{\chi}$  and $\mu_h$ which are not essential for SNR.  Then, we use the fact that SNR requires $n_\chi \mu_\chi/\Lambda$ to be of order one (see Eq.~(\ref{eq:snralign1})), which allows us to substitute $\mu_\chi$ with $\Lambda/n_\chi$. Finally, we will not make a distinction between $H_u$ and $H_d$, assuming that we work in the alignment limit of 2HDM and considering only the light physical Higgs boson which is contained in both $H_{u}$ and $H_{d}$. 
After having  done this, the relevant scalar potential reads (neglecting order-one numerical factors)
\bea\label{eq:vappsimp}
V &=& \left[\frac 1 {n_\chi^2} \Lambda^2 |\chi_i|^2  
+ \frac{1}{n_\chi} |\chi_i|^2 |H|^2 
+ \frac 1 {\Lambda^2} \left\{|H|^2 |\chi_1.\chi_2|^2 + |\chi_i|^2 |H|^4 \right\} 
\right]   \nn \\
& \times & \sum_{k_1, k_2=0}^{\infty} 
\left[\frac{|\chi_i|^2}{\Lambda^2} \right]^{k_1} 
\left[\frac{|H|^2}{\Lambda^2} \right]^{k_2}
. 
\eea
This parametric form of Lagrangian can be obtained explicitly by integrating out fields $S_{1,2}$ with mass $\Lambda$, transforming as $(2,n), (\bar 2,\bar n)$ under ${\rm SU}(2)_L\times {\rm U}(n_\chi)$ with a renormalizable superpotential.

We would now like to estimate the size of various thermal corrections to the Higgs and $\chi$ masses, counting the powers of $n_\chi$ and $T/\Lambda$ (which are expected to be respectively $\gg 1$ and $\ll 1$) that affect the loop series convergence. The powers of $T$ are simply deduced from dimensional analysis. The powers of $n_\chi$, besides coming from the operator coefficients, are generated by the closed $U(n_\chi)$ ``color'' lines. By inspecting various terms in Eq.~(\ref{eq:vappsimp}) we conclude that the leading-loop corrections to the mass operators have the parametric form
\be
\delta V^{(1L)}_T \sim T^2 |H|^2 + \left\{\frac 1 {n_\chi} + \frac {T^2}{\Lambda^2}\right\} T^2 |\chi_i|^2.
\ee
The first term is the one-loop SNR Higgs mass correction derived from the $|\chi_i|^2 |H|^2$ operator in Eq.~(\ref{eq:vappsimp}). The first term in the brackets is generated at one loop from the same operator $|\chi_i|^2 |H|^2$. The second term in the brackets is obtained at two loops from dimension-six operators in the first line of Eq.~(\ref{eq:vappsimp}), and from dimension-two and -four operators dressed respectively with $k_1=2$ and $k_1=1$ powers of $|\chi_i|^2/\Lambda^2$. 
As one can see, the one-loop thermal Higgs mass stays finite in the used large-$n_\chi$ limit, as desired.  

As for the higher-loop effects, we will just state the estimated expansion parameters for the leading loop series. There are two distinct ways of forming the leading loop series; the first manner is by using the same operator dressed with an increasing number of $|\chi_i|^2/\Lambda^2$ factors, leading to the loop expansion parameter $n_\chi T^2/\Lambda^2$.
The second series type is the one formed by multiple insertions of the same operator, in which case the leading effect is produced by the dimension-six operators in the first line of Eq.~(\ref{eq:vappsimp}). The corresponding expansion parameter is also $n_\chi T^2/\Lambda^2$.

As for the fermionic part of the Lagrangian, at two-derivative order it coincides with the one of Eq.~(\ref{eq:lagferm}), with $c_\chi=0$. Higher-order thermal corrections induced by the corresponding dimension-five operators with two fermions were studied in Ref.~\cite{Matsedonskyi:2020mlz}, where it was found that the series' convergence requires $n_\chi T^2/\Lambda^2\ll1$. 

All the higher-loop effects can therefore be suppressed if (up to numerical loop factors) $n_\chi T^2/\Lambda^2\ll 1$, which is the same condition that is needed to suppress the two-loop correction to the Higgs mass computed in Appendix~\ref{sec:twoloop}.

Let us now comment on the dimension-six operator $c_\chi |\chi_i|^2 |\chi_1.\chi_2|^2/\Lambda^2$ which we assumed negligible so far. It turns out that a series of diagrams with multiple insertions of this operator behaves as $(c_\chi n_\chi^2 T^2/ \Lambda^2)^{p}$. Using the requirements $T \gtrsim \mu_\chi$ and $n_\chi \mu_\chi/\Lambda \gtrsim 1$, we see that the series does not converge unless the coefficient $c_\chi$ is suppressed. As we have mentioned earlier, a suppressed value of $c_\chi$ is not a problem for SNR, and also there exist UV completions which do not produce such an operator at tree level.

\section{Two-loop thermal corrections} \label{sec:twoloop}

In this section we present the two-loop corrections to the Higgs mass generated by the dimension-six scalar and dimension-five fermion-scalar interactions. Corresponding diagrams are shown in Figure \ref{fig:2_loop}. The relevant dimension-six operators are
\be
{\cal L} = - \frac{c_{\chi h}^2}{\Lambda^2}  |H_i|^2  |\chi_1.\chi_2|^2 - \frac{c_{\chi h}^2}{\Lambda^2}  |\chi_i|^2 |H_u.H_d|^2.
\ee
The resulting two-loop correction to the Higgs potential is 
\be
\delta V^{(2L,\chi)} = \left\{n_\chi + 2 n_\chi \right\} \frac {c_{\chi h}^2}{\Lambda^2} I_B^2(m_\chi) |H_i|^2, 
\ee
with
\bea
I_B(m) &=& 
  \int \frac{d^3 p}{(2\pi)^3} \frac{1}{2\sqrt{p^2+m^2}}  +\frac{T^2}{2\pi^2} \tilde I_B[m^2/T^2],
\eea
where we are interested in the second part representing the pure thermal correction with
\be
\tilde I_B[x] = \int_0^{\infty} dk \frac{k^2}{\sqrt{k^2+x^2}} \frac{1}{e^{\sqrt{k^2+x}}-1}.
\ee
In the high-$T$ limit we get $\tilde I_B[0] = \pi^2/6$. Overall, we obtain
\be
\delta V^{(2L,\chi)}_T|_{m_i\to 0} \simeq \left\{n_\chi + 2 n_\chi \right\} \left(\frac {c_{\chi h}^2}{\Lambda^2}\right) \left(\frac{T^2}{12}\right)^2 |H_i|^2.
\ee

As for the Higgs-$\tilde \chi$ interactions, they come from
\bea
{\cal L} &=& - \tilde \chi_1.\tilde \chi_2  \left(\mu_\chi + \frac {c_{\chi h}}{\Lambda}H_u.H_d\right) + {\rm h.c.} 
\eea
The two-loop correction to the Higgs potential is (we apply $m_{\tilde \chi},m_H \to 0$ limit from the start)
\be
\delta V^{(2L,\tilde \chi)} = - n_\chi \frac{c_{\chi h}^2}{\Lambda^2} \left( 2 I_B(0) I_F(0) - I_F(0)^2 \right) |H_i|^2,
\ee
with
\bea
I_F(m) 
&=&  \int \frac{d^3 p}{(2\pi)^3} \frac{1}{2\sqrt{p^2+m^2}}  - \frac{T^2}{2\pi^2} \tilde I_F[m^2/T^2],
\eea
where we are interested in the second part representing the pure thermal correction with the function
\be
\tilde I_F[x] = \int_0^{\infty} dk \frac{k^2}{\sqrt{k^2+x^2}} \frac{1}{e^{\sqrt{k^2+x}}+1},
\ee
simplifying at low masses to $\tilde I_F[0] = \pi^2/12$. The final form of the fermionic correction is
\be
\delta V^{(2L,\tilde \chi)}_T|_{m_i\to 0} \simeq  \frac 5 4 n_\chi \left(\frac {c_{\chi h}^2}{\Lambda^2}\right) \left(\frac{T^2}{12}\right)^2 |H_i|^2.
\ee

Overall, the leading two-loop thermal correction is
\be
\delta V^{(2L)}_T|_{m_i\to 0} \simeq \left\{ \frac {17} 4 \frac{1}{144}n_\chi \frac{c_{\chi h}^2 T^2}{\Lambda^2} \right\} T^2 |H_i|^2.
\ee


\bibliographystyle{JHEP}
\bibliography{susy_snr-JHEP2}

\end{document}